\documentclass[prd,onecolumn,showpacs,floatfix,superscriptaddress,nofootinbib]{revtex4-2}
\usepackage[utf8]{inputenc}
\usepackage{dcolumn}
\usepackage{graphicx}
\usepackage{amsmath}
\usepackage{amsfonts}
\usepackage{amssymb}
\usepackage{gensymb}
\usepackage{microtype}
\usepackage{subfigure}
\usepackage{makeidx}
\usepackage{bm}
\usepackage{epsf}
\usepackage{color}
\usepackage{xcolor}
\usepackage{multirow,dcolumn}
\usepackage{graphicx}
\usepackage{mathrsfs}
\usepackage{float}
\usepackage{cancel}
\graphicspath{{Images/}}
\usepackage{geometry}
\usepackage[colorlinks, linkcolor=blue, anchorcolor=blue, citecolor=blue]{hyperref}

\begin{document}

\title{A Photon Cloud Induced from an Axion Cloud} 

\author{Zi-Yu Tang}
\email{tangziyu@ibs.re.kr}
\affiliation{Cosmology, Gravity and Astroparticle Physics Group, Center for Theoretical Physics of the Universe, Institute for Basic Science, Daejeon 34126, Korea}

\author{Eleftherios Papantonopoulos}
\email{lpapa@central.ntua.gr}
\affiliation{Physics Division, School of Applied Mathematical and Physical Sciences, National Technical University of Athens, 15780 Zografou Campus, Athens, Greece.}

\vspace{3.5cm}

\begin{abstract}

It is known that the axion-photon coupling can lead to quantum stimulated emission of photons and classic exponential amplification of electromagnetic (EM) fields  at half the axion mass frequency, when the axion density or the coupling constant is sufficiently large. In this work, we studied the EM photon cloud induced from an axion cloud around a Kerr black hole in the first order of the coupling constant classically. In the presence of a static EM background (such as the extended Wald solution motivated by astrophysical environments), we found that an EM photon cloud emerges, oscillating at the same frequency as the axion cloud and growing exponentially in accordance with the axion cloud when the superradiant condition for the axion field is satisfied. The evolution of the EM photon cloud with time and azimuthal angle is obtained analytically while the cross-sectional distribution is solved numerically. The induced EM field exhibits symmetries that are markedly different from those of the background EM field. Consequently, the induced photon cloud forms an unstable bound configuration that emits EM waves to spatial infinity while being replenished by the axion cloud, providing a potential observational signature of both the presence of an axion cloud and axion-photon coupling. 

\end{abstract}

\maketitle

\section{Introduction}
\label{Introduction}

Axions or axion-like particles were introduced to solve the strong CP problem arising in QCD as it was found a discrepancy between the observed and predicted magnitude of the neutron electric dipole moment  in the strong sector of the standard model~\cite{Peccei:1977hh,Weinberg:1977ma,Wilczek:1977pj}. Also axions appear in supersymmetric theories, and theories with extra dimensions including string theory from the compactification of the extra dimensions~\cite{Arvanitaki:2009fg,Arvanitaki:2010sy,Marsh:2015xka}. In (3+1)-dimensional case, a type called string-model independent axion~\cite{Duncan:1992vz,Svrcek:2006yi} is dual to the field strength of the Kalb-Ramond, spin-one, antisymmetric tensor field, and plays the role of a totally antisymmetric torsion in the geometry. 

Recently, there have been extensive experimental and observational efforts to search for axionic imprints. Superradiant mechanisms provide a promising avenue for probing their existence through gravitational waves~\cite{Yoshino:2013ofa,Arvanitaki:2014wva,Siemonsen:2022yyf}, as axions can form dense clouds around spinning black holes. Although tidal perturbations from a nearby binary companion can destabilize the boson cloud and even terminate superradiance~\cite{Tong:2022bbl}, the binary evolution can be significantly modulated by the backreaction with reduced termination rate~\cite{Fan:2023jjj}. The superrradiance mechanism is a well established process for extracting energy from a rotating black hole~\cite{Brito:2015oca,Brito:2014wla}. Penrose and Floyd~\cite{Penrose:1971uk} first demonstrated that a particle entering the ergoregion of a rotating black hole can split such that one fragment falls in with negative energy, allowing the other to escape with excess energy, thereby extracting the black hole’s rotational energy. Subsequently, Misner~\cite{Misner:1972kx} showed that waves with frequencies $\omega<m\Omega_H$ can also extract rotational energy through superradiant scattering, in which an incoming wave is amplified as it scatters off a rotating black hole. Later, Press and Teukolsky \cite{Press:1972zz} introduced the concept of the black hole bomb, and Teukolsky further generalized the mechanism to EM and gravitational waves~\cite{Teukolsky:1973ha}. Nevertheless, the Kerr solution remains stable against massless scalar, EM and gravitational perturbations without reflection.

It is well known that an isolated black hole cannot sustain an intrinsic EM field unless it carries a net electric charge. Thus, for a black hole to exhibit EM effects, some charging mechanism is required. In realistic astrophysical environments, surrounding plasma and accretion flows in the presence of large-scale magnetic fields can give rise to effective charge separation or induce a nonvanishing net charge on the black hole. Possible origins of such EM configurations include charge separation induced by magnetospheric processes such as the Blandford–Znajek mechanism, effects associated with plasma inflow, disk–magnetosphere interactions, as well as pair-production–supported magnetospheres. In the present work, however, we do not model these mechanisms explicitly and instead adopt the extended Wald solution \cite{Wald:1974np} as an effective, stationary background description. Neglecting backreaction and making use of the two Killing vectors of the Kerr spacetime, Wald constructed an exact test solution of Maxwell's equations describing a homogeneous magnetic field aligned with the black hole spin axis, which could lead to the accretion of charges from the surrounding plasma in the accretion disk or the interstellar medium. As a result, the black hole would acquire a charge and be described by a Kerr-Newman spacetime. However, at equilibrium with the original magnetic field, this EM configuration is also a source-free test solution on the Kerr metric, so that we can consider it as the EM background solution in our work. Besides, extraction of energy and charge from a black hole was studied in~\cite{Bekenstein:1973mi}. The superradiance effects and the multimessenger signals of an EM field were studied in~\cite{Siemonsen:2022ivj,Siemonsen:2022yyf}. Recently, new polarized images from the Event Horizon Telescope (EHT) collaboration revealed significant variations in brightness and polarization patterns across 2017, 2018, and 2021, highlighting the unexpectedly evolving magnetic field structures around the black hole M87*~\cite{EventHorizonTelescope:2025vum}.

Although the axion-photon coupling is expected to be very weak, it was suggested that the quantum stimulated emission of photons can be generated at large enough axion number~\cite{Kephart:1986vc,Kephart:1994uy}, and conjectured that blasts of light could be emitted from black hole systems~\cite{Rosa:2017ury,Sen:2018cjt}. Recently at classical level, it was shown that EM fields can be exponentially amplified in Kerr-axion cloud system, when the amplitude of the axion cloud multiplied by the coupling constant is larger than a critical value~\cite{Ikeda:2018nhb}. Thereinto, a fraction of the cloud energy is transferred to the EM blasts, while the depleted axion cloud will be replenished via superradiance, leading to the spin-down of the black hole as it was found in~\cite{Kuang:2024ugn}.  In~\cite{Chatzifotis:2022ene}, it was also found that the presence of an axion field can cause the reversal of rotation in Kerr black holes. Even in the presence of plasma, the EM instability is still controlled by the axionic coupling~\cite{Spieksma:2023vwl}. Here the plasma effects are characterized by the plasma frequency $\omega_p$, which acts as an effective mass scale for EM waves and governs plasma screening. Interestingly, a new class of superradiant instabilities of axion bound states in neutron star magnetospheres has been identified \cite{Day:2019bbh}, in which axion-photon mixing allows the extraction of rotational energy from the neutron star, since rotation and a dissipation channel are enough to trigger superradiance \cite{Brito:2015oca}.

In a flat-spacetime analysis with a uniform oscillating axion background, the instability appears through the Mathieu equation, originating from the homogeneous part of the equation \cite{Boskovic:2018lkj}. In contrast, the flat-spacetime analysis in a homogeneous EM background showed that the axion-photon coupling modifies the dispersion relation, and that a sufficiently strong electric field can trigger an instability with a growing mode. This latter instability arises from the inhomogeneous (sourced) part of the EM field equation. In fact, in the presence of external EM fields, photons can mix with low-mass bosons if they couple via a two-photon vertex, analogous to neutrino-flavor-mixing effects~\cite{Raffelt:1987im}. In addition, both axion-photon mixing around charged black holes and dark photon-photon mixing around neutral black holes were studied in the presence of plasma~\cite{Cannizzaro:2024hdg}, in which the flat spacetime analysis in the presence of a magnetic field gave the mixing matrixes with non-diagonal terms. Moreover, superradiant instabilities by photons confined around black holes in the 
presence of astrophysical plasmas were investigated under the plasma-photon interaction, where quasi-bound EM states were found to be superradiantly unstable for rotating black holes though it probably be quenched due to plasma blowout in the nonlinear evolution~\cite{Cannizzaro:2020uap,Cannizzaro:2021zbp,Cannizzaro:2023ltu}. From the observational side, with the polarimetric measurements of M87* by the EHT collaboration stringent constraints on the axion-photon coupling constant in the mass range of $\sim\left(10^{-21}-10^{-20}\right)$eV were given, based on the azimuthal distribution of the electric vector position angle~\cite{Chen:2021lvo}. Furthermore, the deviation of photon geodesics due to oscillating metric perturbations induced from the superradiant clouds was predicted in~\cite{Chen:2022kzv}.

In this work we studied the EM photon cloud induced from an axion cloud via the axion-photon coupling surrounding a Kerr black hole. In the first order of the coupling constant, we found that a background EM field is necessary to have a first-order induced EM structure, and if it is stationary then the frequency of the induced photon cloud will be the same as the frequency of the axion cloud. It means that it grows or decays in accordance with the axion cloud when the superradiant condition for the axion field is satisfied or not, while the oscillating frequency is also the same. It is well known that the timescale of the superradiant instability is much larger than the timescale of the oscillation, therefore we can ignore the growth of the amplitude and focus on the oscillating part.

 We found  that no threshold on the coupling constant or on the amplitude of the axion cloud is required for the emergence of this oscillating and growing photon cloud. The instability is inherited from the axion cloud in the presence of an EM background. This is in contrast to the case without a background EM field, where a threshold for the amplitude of the axion cloud multiplied by the coupling constant and an initial EM pulse are both required; otherwise, no instability develops and the initial EM fluctuation decays exponentially. While for the dark photon-photon coupling case, a dark photon cloud can transfer its energy to form a rotating EM field directly without any external EM fields~\cite{Siemonsen:2022ivj}, since no background EM fields are necessary for the mixing to occur~\cite{Cannizzaro:2024hdg}.

The work is organized as follows. Firstly in Sec. \ref{First-order Analysis of the Axion-Photon Coupling} we study the basic field equations in the first-order of the coupling constant, then in Sec. \ref{A Photon Cloud Induced from an axion cloud} we show how an EM photon cloud is induced from the axion cloud in the presence of an EM background and discuss symmetries. Moreover, in Sec. \ref{The induced electric and magnetic fields} we study the induced electric and magnetic fields respectively.  Finally in Sec. \ref{Conclusions} we conclude.

\section{First-order Analysis of the Axion-Photon Coupling}
\label{First-order Analysis of the Axion-Photon Coupling}

In this work, we consider the interaction between an axion cloud and the EM field around a Kerr black hole with the action 
\begin{eqnarray}
  && S=\int d^4 x \sqrt{-g}\left[\frac{R}{2\kappa}+\mathcal{L}_m\right]~, \\
  && \mathcal{L}_m=-\frac{1}{4}F_{\mu\nu}F^{\mu\nu}-\frac{1}{2}\nabla_\mu \phi \nabla^\mu \phi -\frac{1}{2}\mu^2\phi^2-\frac{k_a}{2}\phi F_{\mu\nu}{}^*F^{\mu\nu}~, 
\end{eqnarray}
where the axion field $\phi$ is a real pseudo scalar field with mass $m_a\equiv \mu \hbar$, $k_a$ is the coupling constant, $F_{\mu\nu}\equiv \nabla_\mu A_\nu -\nabla_\nu A_\mu$ is the electromagnetic field strength tensor and ${}^*F^{\mu\nu}\equiv \frac{1}{2}\epsilon^{\mu\nu\rho\sigma}F_{\rho\sigma}$ is the dual field strength tensor. Here $\epsilon^{\mu\nu\rho\sigma}\equiv \frac{1}{\sqrt{-g}}E^{\mu\nu\rho\sigma}$ and $E^{\mu\nu\rho\sigma}$ denotes the totally antisymmetric Levi-Civita symbol with $E^{0123}=+1$. We use geometrized units $c=G=1$ unless otherwise specified. 

By variation of the above action, we  obtain the basic field equations
\begin{eqnarray}
&& \nabla^\mu \nabla_\mu \phi -\mu^2\phi-\frac{k_a}{2}F_{\mu\nu}{}^*F^{\mu\nu}=0~, \label{eqKG}\\
&& \nabla_\nu F^{\mu\nu}+2k_a \nabla_\nu \phi {}^*F^{\mu\nu}=0~, \label{eqF}\\
&& R_{\mu\nu}-\frac{1}{2}g_{\mu\nu} R=\kappa \left(g_{\mu\nu}\mathcal{L}_m-2\frac{\delta \mathcal{L}_m}{\delta g^{\mu\nu}}\right)\equiv \kappa T_{\mu\nu}~, 
\end{eqnarray}
with 
\begin{eqnarray}
    \frac{\delta \mathcal{L}_m}{\delta g^{\mu\nu}}=-\frac{1}{2}F^\sigma {}_\mu F_{\sigma\nu}-\frac{1}{2}\nabla_\mu \phi \nabla_\nu \phi-2k_a \phi F^\sigma{}_\nu {}^*F_{\sigma\mu}~.
\end{eqnarray}

In our study, the Kerr geometry is fixed as the background, since the backreactions of both axion cloud and the EMs on the metric are in the second-order of the fluctuations.  To clarify the hierarchy, we introduce small quantities $\eta$ and $e$ to represent the magnitudes of the axion cloud and the EM field respectively as $F_{\mu\nu}\to e F_{\mu\nu}$ and $\phi \to \eta \phi$, and the energy-momentum tensor becomes
\begin{eqnarray}
    T_{\mu\nu}&\to&-\frac{1}{2}g_{\mu\nu}\left(\frac{1}{2}e^2F_{\alpha\beta}F^{\alpha\beta}+\eta^2\nabla_\sigma \phi \nabla^\sigma \phi +\mu^2\eta^2\phi^2+k_a e^2\eta\phi F_{\alpha\beta}{}^*F^{\alpha\beta}\right)~\notag\\
    &&+e^2F^\sigma {}_\mu F_{\sigma\nu}+\eta^2\nabla_\mu \phi \nabla_\nu \phi+4k_a e^2\eta\phi F^\sigma{}_\nu {}^*F_{\sigma\mu}~, 
\end{eqnarray}
showing that the magnitudes of the backreaction on the metric are of the orders $\sim e^2,~\eta^2,~\mu^2\eta^2,~k_a \eta e^2$. The other field equations become
\begin{eqnarray}
    && \nabla^\mu \nabla_\mu \phi -\mu^2\phi-\frac{k_a e^2}{2\eta}F_{\mu\nu}{}^*F^{\mu\nu}=0~, \\
&& \nabla_\nu F^{\mu\nu}+2k_a \eta\nabla_\nu \phi {}^*F^{\mu\nu}=0~, \\
\end{eqnarray}
which yields the conditions $\mu^2 \eta \gg k_a e^2$ and $k_a\eta \ll 1$ when we consider the axion cloud and the EM field as background fields, with the effects of their interaction appearing only at higher orders. 

Using the typical field amplitude in the boson cloud $\langle \phi^2\rangle\sim 8 \tilde{a} \alpha^5 M_{\rm Pl}^2$ \cite{Fukuda:2019ewf}, and the limit for the largest magnetic field $B\sim 4\times 10^{18}\left(M/M_\odot\right)^{-1/2}{\rm Gauss}$ that can be supported by an accretion disk around astrophysical BHs \cite{Rees:1984si}, we obtain the first condition 
\begin{equation}
    k_a \ll \frac{160.705 \sqrt{\tilde{a}}\alpha^{9/2}}{M_{\rm BH}/M_\odot}~{\rm GeV}^{-1}~
\end{equation}
that the background solution of the axion cloud without EM field is valid, and the  relevant axion mass is around
\begin{equation}
    m_a\sim \frac{\alpha}{M_{\rm BH}/M_\odot}1.3\times 10^{10}~{\rm eV}~.
\end{equation}
Here we adopt natural units, setting $\hbar=c=1$ and $4\pi\epsilon_0=1$, in order to facilitate comparison with other constraints on the coupling constant. For instance, in the case of the supermassive black hole M87*, taking $\tilde{a}=0.8$ and $\alpha=0.2$, we have $k_a \ll 1.58233\times 10^{-11}~{\rm GeV}^{-1}$. In contrast, for a primordial black hole with $M/M_\odot=0.2$, $\tilde{a}=0.5$ and $\alpha=0.2$, the condition becomes much less restrictive, $k_a \ll 0.406555~{\rm GeV}^{-1}$. We will verify the second validity condition of the perturbative framework for the EM field equation, after we obtain the amplitude of the first-order EM field. 

We now return to the original field equations and expand them to first order in the coupling constant $k_a$ 
\begin{eqnarray}
    &&\nabla^\mu \nabla_\mu \phi_{(0)} -\mu^2\phi_{(0)}+k_a\left(\nabla^\mu \nabla_\mu \phi_{(1)} - \mu^2\phi_{(1)}-\frac{1}{2}F_{\mu\nu}^{(0)}{}^*F^{\mu\nu}_{(0)}\right)+\mathcal{O}\left(k_a^2\right)=0~,\\
    &&\nabla_\nu F^{\mu\nu}_{(0)}+k_a \left(\nabla_\nu F^{\mu\nu}_{(1)}+2\nabla_\nu \phi_{(0)} {}^*F^{\mu\nu}_{(0)}\right)+\mathcal{O}\left(k_a^2\right)=0~, \label{eqF1}
\end{eqnarray}
which clearly shows that to zeroth order the axion cloud and the EM field obey their own field equations respectively, while at first order the background EM field gives a substructure on the axion cloud and the coexistence of the axion cloud and the background EM field generates an extra EM field of order $k_a$. The first-order backreaction on the axion cloud has been studied in~\cite{Boskovic:2018lkj}. In this work, we focus on the EM field induced from the axion cloud in the presence of a background EM field.

The Kerr metric in Boyer-Lindquist (BL) coordinates is
\begin{equation}
    ds^2=-\frac{1}{\Sigma}\left(\Delta_r-a^2\sin^2{\theta}\right)dt^2+
    \frac{\Sigma}{\Delta_r}dr^2+\Sigma d\theta^2+\frac{\Gamma}{\Sigma}\sin^2{\theta}d\varphi^2-\frac{4a M r}{\Sigma}\sin^2{\theta}dtd\varphi~, \label{metric}
\end{equation}
where 
\begin{eqnarray}
   &&\Sigma=r^2+a^2\cos^2{\theta}~, \\
   &&\Delta_r=r^2-2Mr+a^2~, \\
   &&\Gamma =\left(a^2+r^2\right)^2- \Delta_r a^2\sin ^2{\theta}~.
\end{eqnarray}
For the background EM field we consider the extended Wald solution 
\begin{equation}
    A_\mu^{(0)}=A_\mu^{\rm Wald}=\frac{B_0 \sin^2{\theta}}{2\Sigma}\left(-2aM r,0,0,\Gamma\right)~, \quad A^\mu_{(0)}=A^\mu_{\rm Wald}=\left(0,0,0,\frac{B_0}{2}\right)~,
\end{equation}
which describes the equilibrium state of the black hole absorbing charges and discharging processes from the surroundings of astrophysical environments such as the accretion disk, plasma and the interstellar medium~\cite{Wald:1974np}. It is worth noting that this extended Wald solution is still a source-free test solution on the Kerr geometry.

For the axion cloud part, the bound states in Kerr metric can be obtained analytically in approximation when the fine structure constant $\alpha \equiv GM\mu/c=M\mu \ll 1$ in separable form as 
\begin{equation}
    \phi_{(0)}=e^{-i\omega t+im\varphi}S_{lm}(\theta)R_{lm}(r)~ \label{phi}
\end{equation}
with $\omega=\omega_R+ i \omega_I $ that~\cite{Detweiler:1980uk}
\begin{eqnarray}
    \omega_R &\simeq &\mu \left(1-\frac{\alpha^2}{2\left(n+l+1\right)^2}\right)~, \\
    \omega_I &\simeq & \frac{1}{\gamma_{nlm}M}\left(\frac{a m}{M}-2\mu r_+\right)\alpha^{4l+5}
\end{eqnarray}
where $\gamma_{nlm}$ is a constant, $n$ is the principal quantum number, and $l$ corresponds to the orbital angular momentum quantum number. The condition for the superradiance to occur $\mu<\frac{am}{2M r_+}=m \Omega_H$ is equivalent to $\alpha<\frac{am}{2r_+}=\frac{ma}{2M\left(1+\sqrt{1-a^2/M^2}\right)}$. 

As we can see in eq. (\ref{eqF1}), since the background EM field $F_{\mu\nu}^{(0)}$ is independent of time $t$ and azimuthal angle $\varphi$, therefore the induced first-order EM field in form $F^{\mu\nu}_{(1)} \sim e^{-i\omega_1 t+im_1\varphi}$ must have same frequency as the axion field (\ref{phi}) that $\omega_1=\omega$ and $m_1=m$. It means that the induced EM field grows or decays in accordance with the axion cloud when the superradiant condition for the axion field is satisfied or not, while the oscillating frequency is also the same. It is well known that the timescale of the superradiant instability is much larger than the timescale of the oscillation, therefore we can ignore the growth of the amplitude and focus on the oscillating part.

It is necessary to clarify the difference in the EM frequencies. Returning to the original Klein-Gordon equation (\ref{eqKG}), an axion field evolving as $\phi \sim e^{-i\omega t+im\varphi}$ would lead to the EM field as $F_{\mu\nu} \sim e^{-i\omega_{\rm EM} t+im_{\rm EM}\varphi}$ at half frequency of the axion field $\omega_{\rm EM}=\omega/2$. However, if we look at the EM field equation (\ref{eqF}), a direct relation between the axion and the EM frequencies cannot be obtained. 

In the quantum stimulated case~\cite{Kephart:1986vc}, the half frequency was explained by the axion decay channel $\phi \to 2\gamma$, where each photon carries energy $E=m_a/2$. Numerical simulations~\cite{Ikeda:2018nhb,Spieksma:2023vwl} also showed that the EM oscillation frequency appears at half the axion mass frequency for supercritical couplings. In~\cite{Boskovic:2018lkj}, the origin of the instability was analyzed in flat spacetime through the Mathieu equation, which predicts instabilities for $\omega_{\rm EM}=\mu/2,\mu,3\mu/2,...$. Whereas in~\cite{Spieksma:2023vwl} the frequency peaks at $\mu,2\mu,...$ were absent, and for high axionic couplings in dense plasmas with plasma frequency $\omega_p>\mu/2$, the EM spectrum instead centers at $\omega_{\rm EM}=\mu$ rather than at the usual $\omega_{\rm EM}=\mu/2$. This behavior suggests that the standard axion decay channel into two photons is kinematically forbidden and changes to process $\phi+\phi \to \gamma+\gamma$.

\section{A Photon Cloud Induced from an axion cloud}
\label{A Photon Cloud Induced from an axion cloud}

In this Section we will solve the first-order induced EM field $F^{\mu\nu}_{(1)}$ in the Newman-Penrose (NP) formalism, in which using the Kinnersley's null tetrad~\cite{Kinnersley:1969zza}
\begin{equation}
    l^\mu=\left(\frac{r^2+a^2}{\Delta_r},1,0,\frac{a}{\Delta_r}\right)~,~n^\mu=\frac{1}{2\Sigma}\left(r^2+a^2,-\Delta_r,0,a\right)~,~m^\mu=\frac{1}{\sqrt{2}\left(r+i a \cos{\theta}\right)}\left(i a \sin{\theta},0,1,\frac{i}{\sin{\theta}}\right)~,
\end{equation}
we can construct three NP components 
\begin{equation}
    \varphi_0=F_{\mu\nu}l^\mu m^\nu~, \quad \varphi_1=\frac{1}{2}F_{\mu\nu}\left(l^\mu n^\nu+\overline{m}^\mu m^\nu\right)~, \quad \varphi_2=F_{\mu\nu}\overline{m}^\mu n^\nu~, 
\end{equation}
where $\varphi_0$ is the ingoing radiation with spin weight $s=1$ and $\varphi_2$ represents outgoing radiation with spin weight $s=-1$. Note that there are only two real degree of freedoms though the three components are complex. The EM tensor can be reconstructed as 
\begin{equation}
    F_{\mu\nu}=2\varphi_0 \overline{m}_{[\mu}n_{\nu ]}+2\overline{\varphi}_0 m_{[ \mu}n_{\nu ]}+2\left(\varphi_1+\overline{\varphi}_1\right)n_{[ \mu}l_{\nu ]}+2\left(\varphi_1-\overline{\varphi}_1\right)m_{[ \mu}\overline{m}_{\nu ]}+2\varphi_2 l_{[\mu}m_{\nu ]}+2\overline{\varphi}_2 l_{[\mu}\overline{m}_{\nu ]}~. \label{ReconstructF}
\end{equation}

We derived the perturbation equation for electromagnetic field with sources in NP formalism~\cite{Teukolsky:1973ha}
\begin{eqnarray}
    &&\left(D-2\rho\right)\varphi_1-\left(\bar{\delta}+\pi-2\alpha\right)\varphi_0+\kappa \varphi_2=-\frac{1}{2}J_k~,\\
    &&\left(\delta-2\tau\right)\varphi_1-\left(\Delta+\mu-2\gamma\right)\varphi_0+\sigma \varphi_2=-\frac{1}{2}J_m~,\\
    &&\left(D-\rho+2\varepsilon\right)\varphi_2-\left(\bar{\delta}+2\pi\right)\varphi_1+\lambda \varphi_0=-\frac{1}{2}J_{\bar{m}}~,\\
    && \left(\delta-\tau+2\beta\right)\varphi_2-\left(\Delta+2\mu\right)\varphi_1+\nu \varphi_0=-\frac{1}{2}J_{\bar{m}}~,
\end{eqnarray}
and obtained the decoupled equation for the NP scalar $\varphi_1$ 
\begin{equation}
    \left[\Delta D +\left(\bar{\mu}-\mu-\gamma-\bar{\gamma}\right)D-\bar{\delta}\delta+\left(\bar{\tau}+2\alpha\right)\delta \right]\frac{\varphi_1}{\rho^2}=\left[\Delta+\left(\bar{\mu}-\mu-\gamma-\bar{\gamma}\right)\right]\frac{-J_l}{2\rho^2}+\left[\bar{\delta}-\left(\bar{\tau}+2\alpha\right)\right]\frac{-J_m}{2\rho^2}~,
\end{equation}
where $D\equiv l^a\nabla_a,\Delta\equiv n^a\nabla_a,\delta\equiv m^a\nabla_a, \bar{\delta}\equiv \overline{m}^a\nabla_a$ are derivative operators and $\kappa, \rho,\sigma,\tau,\nu,\mu,\lambda,\pi,\varepsilon,\beta,\gamma,\alpha$ are the spin coefficients. With the Kerr metric in BL coordinates (\ref{metric}), it can be written as
\begin{eqnarray}
    &&\left[\frac{\left(r^2+a^2\right)^2}{\Delta_r}-a^2\sin^2{\theta}\right]\frac{\partial^2 \psi}{\partial t^2}+\frac{4Mar}{\Delta_r}\frac{\partial^2 \psi}{\partial t \partial \varphi}+\left[\frac{a^2}{\Delta_r}-\frac{1}{\sin^2{\theta}}\right]\frac{\partial^2 \psi}{\partial \varphi^2}-\frac{\partial}{\partial r}\left(\Delta_r \frac{\partial \psi}{\partial r}\right)-\frac{1}{\sin{\theta}}\frac{\partial}{\partial \theta}\left(\sin{\theta}\frac{\partial \psi}{\partial \theta}\right)~\notag \\
    &&+\frac{2M\rho^2}{\bar{\rho}}\psi=\frac{1}{\bar{\rho}}\left\{\left[\Delta+\left(\bar{\mu}-\mu-\gamma-\bar{\gamma}\right)\right]\frac{J_l}{\rho^2}+\left[\bar{\delta}-\left(\bar{\tau}+2\alpha\right)\right]\frac{J_m}{\rho^2}\right\}~,
\end{eqnarray}
where $\psi=\varphi_1/\rho$, $J_l=J_\mu l^\mu$ and $J_m=J_\mu m^\mu$. The other two decoupled equations for $\varphi_0$ and $\varphi_2$ with sources can be found in~\cite{Teukolsky:1972my}.

In fact, in our framework, the axion cloud is the source for the induced EM field 
\begin{equation}
    \nabla_\nu F^{\mu\nu}_{(1)}=-2\nabla_\nu \phi_{(0)} {}^*F^{\mu\nu}_{(0)}\equiv J^\mu_{(1)}~.\label{eqFJ}
\end{equation}
We can take a good approximation  to the bound states of axion cloud around a rotating black hole~\cite{Brito:2014wla}
\begin{equation}
    \phi_{(0)}=\alpha^2 \phi_0\frac{ r }{ M}e^{-\frac{ r}{2M}\alpha^2}\cos{\left(\varphi-\omega_R t\right)}\sin{\theta}~,\label{axioncloud}
\end{equation}
which is valid for $\alpha\lesssim 0.2$ even at large black hole spin, and $\omega_R \simeq \mu$ is the oscillating frequency. 
It can be separated into two parts 
\begin{equation}
    \phi_1=e^{-i \omega t+i \varphi}R(r)S(\theta)~,\quad \phi_2=e^{i \omega t-i \varphi}R(r)S(\theta)~,
\end{equation}
with $R(r)=\alpha^2 \phi_0\frac{ r }{ 2M}e^{-\frac{ r}{2M}\alpha^2}$ and $S(\theta)=\sin{\theta}$.

Using $\phi_1$  as the source and the NP scalars as partly separable variables
\begin{equation}
    \varphi_{01}=e^{-i \omega_1 t+i m_1\varphi}f_{01}\left(r,\theta\right)~,~\varphi_{11}=e^{-i \omega_1 t+i m_1\varphi}f_{11}\left(r,\theta\right)~,~\varphi_{21}=e^{-i \omega_1 t+i m_1\varphi}f_{21}\left(r,\theta\right)~,
\end{equation}
we can obtain the three decoupled differential equations respectively for $f_{01}\left(r,\theta\right)$, $f_{11}\left(r,\theta\right)$ and $f_{21}\left(r,\theta\right)$ after dropping off the time $t$ and azimuthal angle $\varphi$ dependence by $\omega_1=\omega$ and $m_1=m=1$. We can also do that for $\phi_2$. Then we replace $\omega=\mu=\alpha/M$, $a_0\equiv a/M$, $r_0 \equiv r/M$, and  define new dimensionless quantities $f_{01M}\equiv M f_{01}\left(r,\theta\right)$, $f_{11M}\equiv M f_{11}\left(r,\theta\right)$, $f_{\rm 21M}\equiv M f_{21}\left(r,\theta\right)$. After absorbing some coefficients from the source part into the coupling constant $k_a$, we now use a new dimensionless coupling constant $\varepsilon \equiv \alpha^2 B_0 M \phi_0 k_a$ as the small quantity for expansion, then the new field equations are dimensionless
\begin{eqnarray}
    &&{\rm eq}f_{\rm 01M}: \left\{ \frac{\partial^2 f_{\rm 01M}\left(r_0,\theta\right)}{\partial r_0^2}, \frac{\partial f_{\rm 01M}\left(r_0,\theta\right)}{\partial r_0},\frac{\partial^2 f_{\rm 01M}\left(r_0,\theta\right)}{\partial \theta^2},\frac{\partial f_{\rm 01M}\left(r_0,\theta\right)}{\partial \theta},f_{\rm 01M}\left(r_0,\theta\right), a_0, \alpha\right\}~, \\
    &&{\rm eq}f_{\rm 11M}: \left\{ \frac{\partial^2 f_{\rm 11M}\left(r_0,\theta\right)}{\partial r_0^2}, \frac{\partial f_{\rm 11M}\left(r_0,\theta\right)}{\partial r_0},\frac{\partial^2 f_{\rm 11M}\left(r_0,\theta\right)}{\partial \theta^2},\frac{\partial f_{\rm 11M}\left(r_0,\theta\right)}{\partial \theta},f_{\rm 11M}\left(r_0,\theta\right), a_0, \alpha\right\}~, \\
    &&{\rm eq}f_{\rm 21M}: \left\{ \frac{\partial^2 f_{\rm 21M}\left(r_0,\theta\right)}{\partial r_0^2}, \frac{\partial f_{\rm 21M}\left(r_0,\theta\right)}{\partial r_0},\frac{\partial^2 f_{\rm 21M}\left(r_0,\theta\right)}{\partial \theta^2},\frac{\partial f_{\rm 21M}\left(r_0,\theta\right)}{\partial \theta},f_{\rm 21M}\left(r_0,\theta\right), a_0, \alpha\right\}~,
\end{eqnarray}
and there are only two dimensionless parameters $a_0$ and $\alpha$. 

Different from the quasi-normal modes and bound states, here the frequency of the induced EM field is already fixed by the frequency of the source (axion cloud), it is not an eigenvalue problem and no boundary condition should be imposed. Actually we can obtain the asymptotic behaviors of the functions at the boundaries by expanding the equations (we choose decaying solutions at spatial infinity), and then solve the equations numerically. The asymptotic behaviors of $f_{01M}$, $f_{11M}$ and $f_{\rm 21M}$ at the boundaries can be found in Appendix \ref{Asymptotic behaviors}. Similarly, for the NP scalars induced from $\phi_2$
\begin{equation}
    \varphi_{02}=e^{i \omega_1 t-i m_1\varphi}f_{02}\left(r,\theta\right)~,~\varphi_{12}=e^{i \omega_1 t-i m_1\varphi}f_{12}\left(r,\theta\right)~,~\varphi_{22}=e^{i \omega_1 t-i m_1\varphi}f_{22}\left(r,\theta\right)~,
\end{equation}
we can get the corresponding equations and boundary conditions for $f_{02M}\left(r_0,\theta\right)$, $f_{12M}\left(r_0,\theta\right)$, $f_{22M}\left(r_0,\theta\right)$. 

\begin{figure}[h]
\centering
  \includegraphics[width=0.3\textwidth]{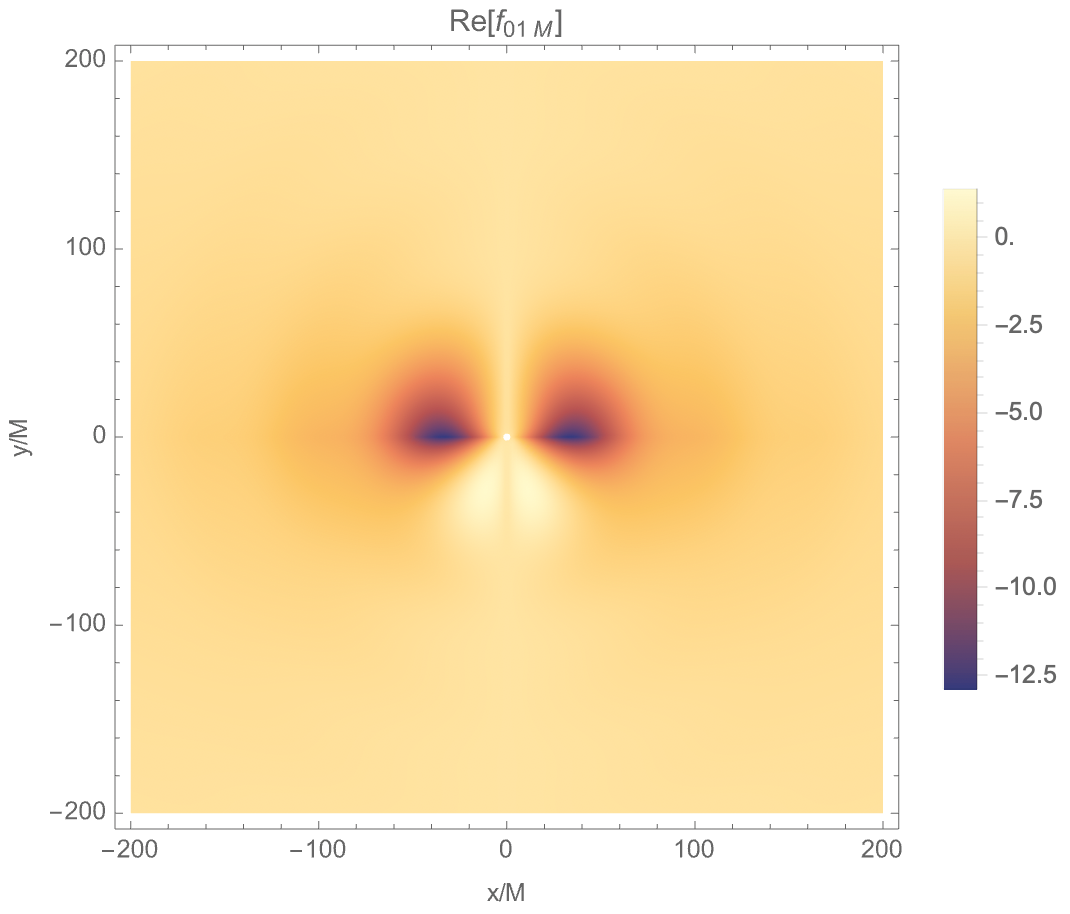}
  \includegraphics[width=0.3\textwidth]{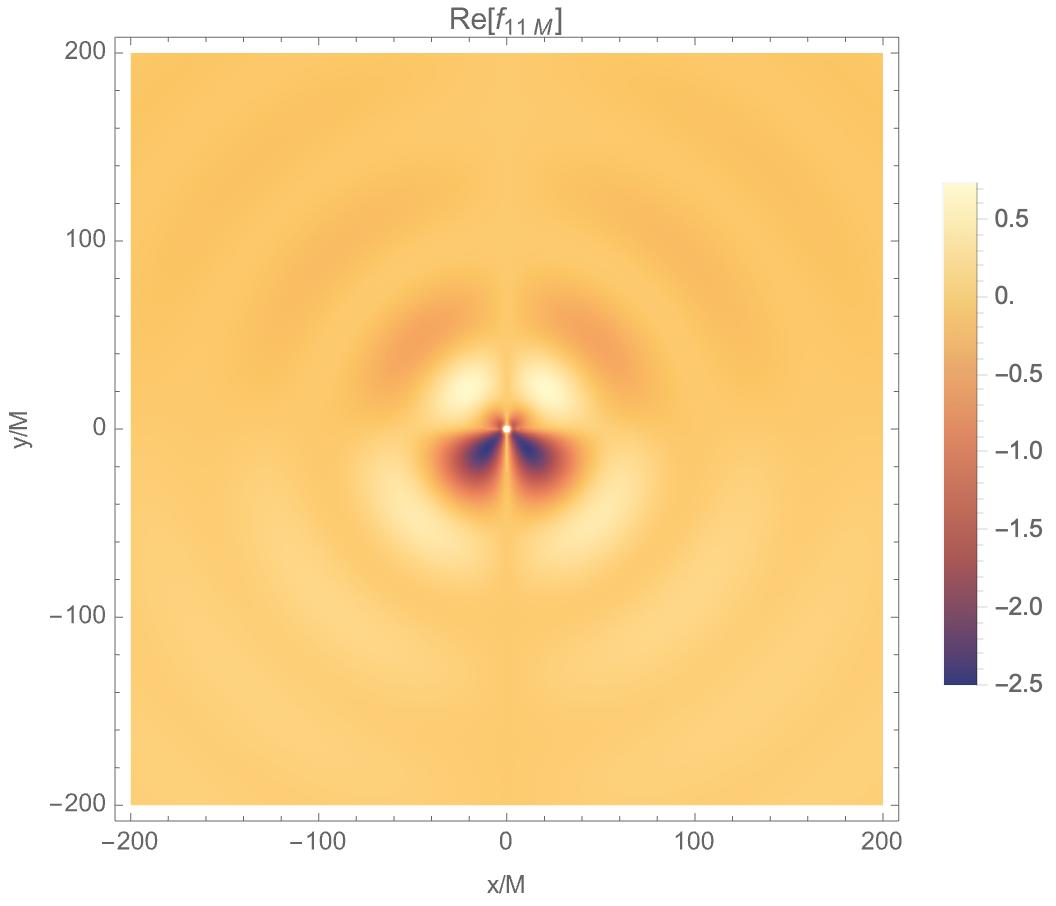}
  \includegraphics[width=0.3\textwidth]{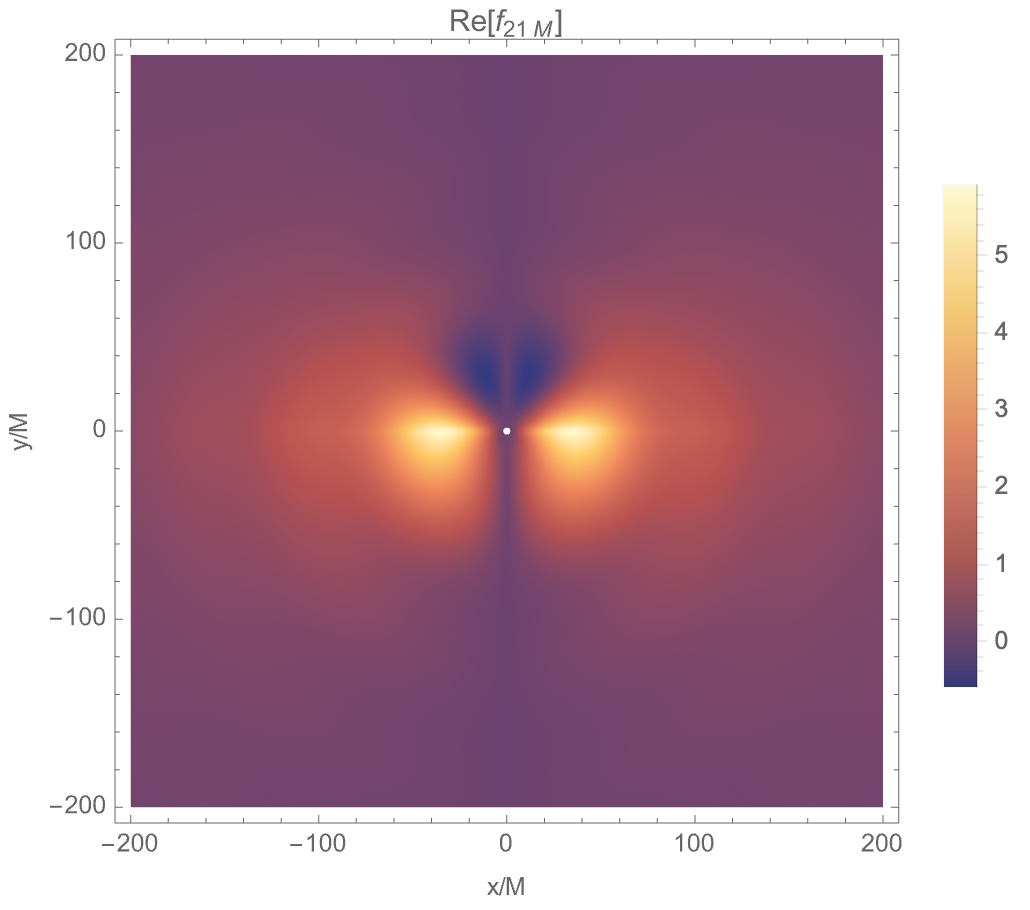}
  \includegraphics[width=0.3\textwidth]{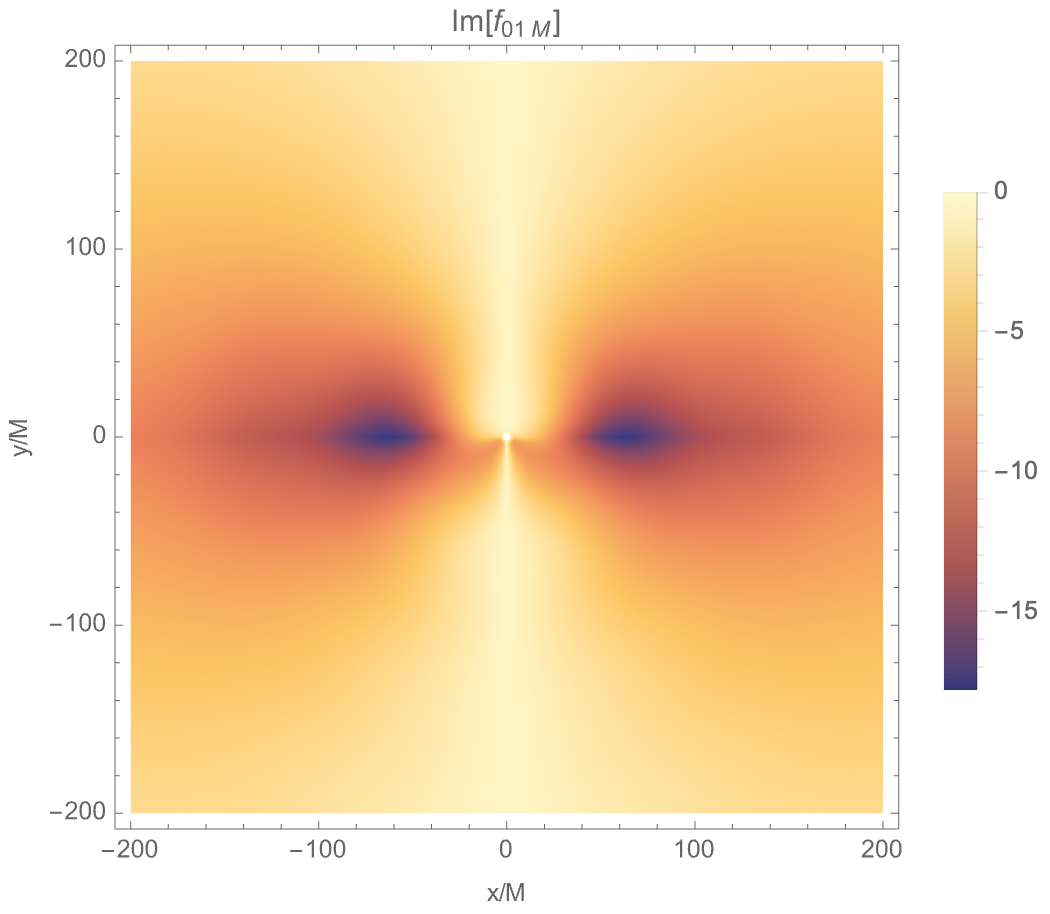}
  \includegraphics[width=0.3\textwidth]{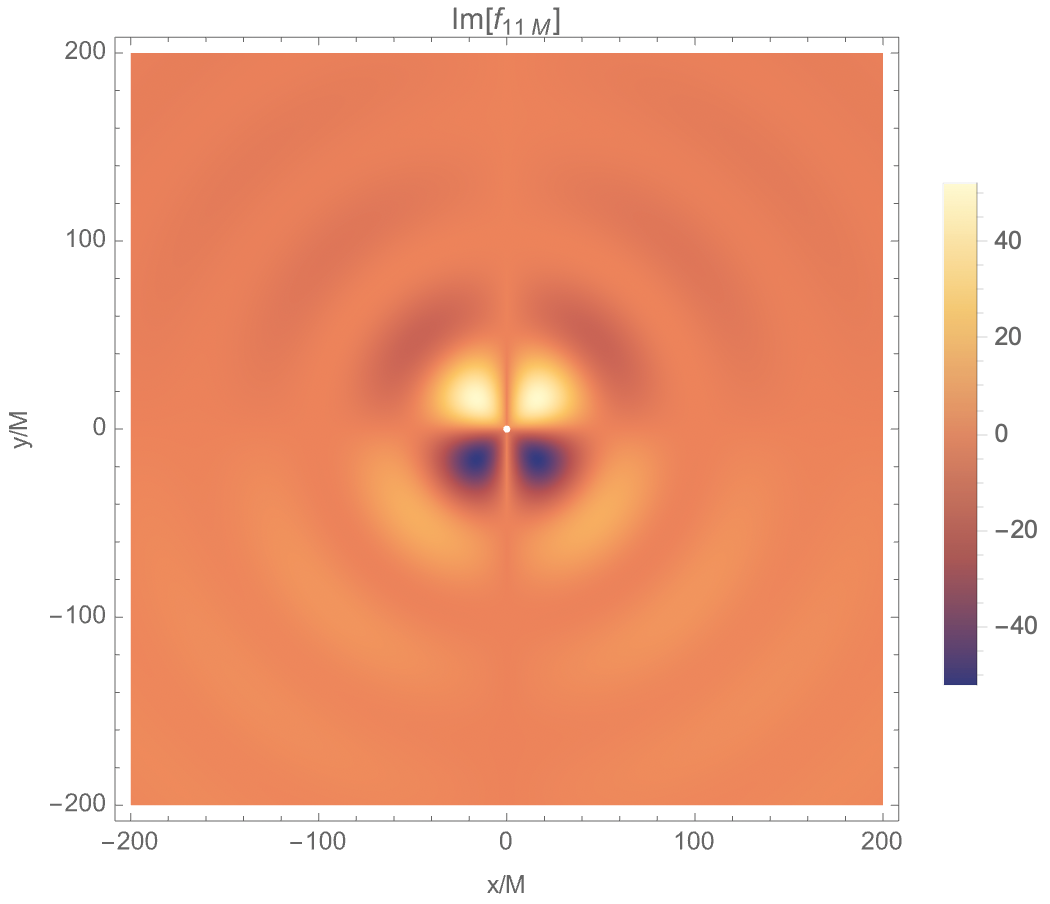}
  \includegraphics[width=0.3\textwidth]{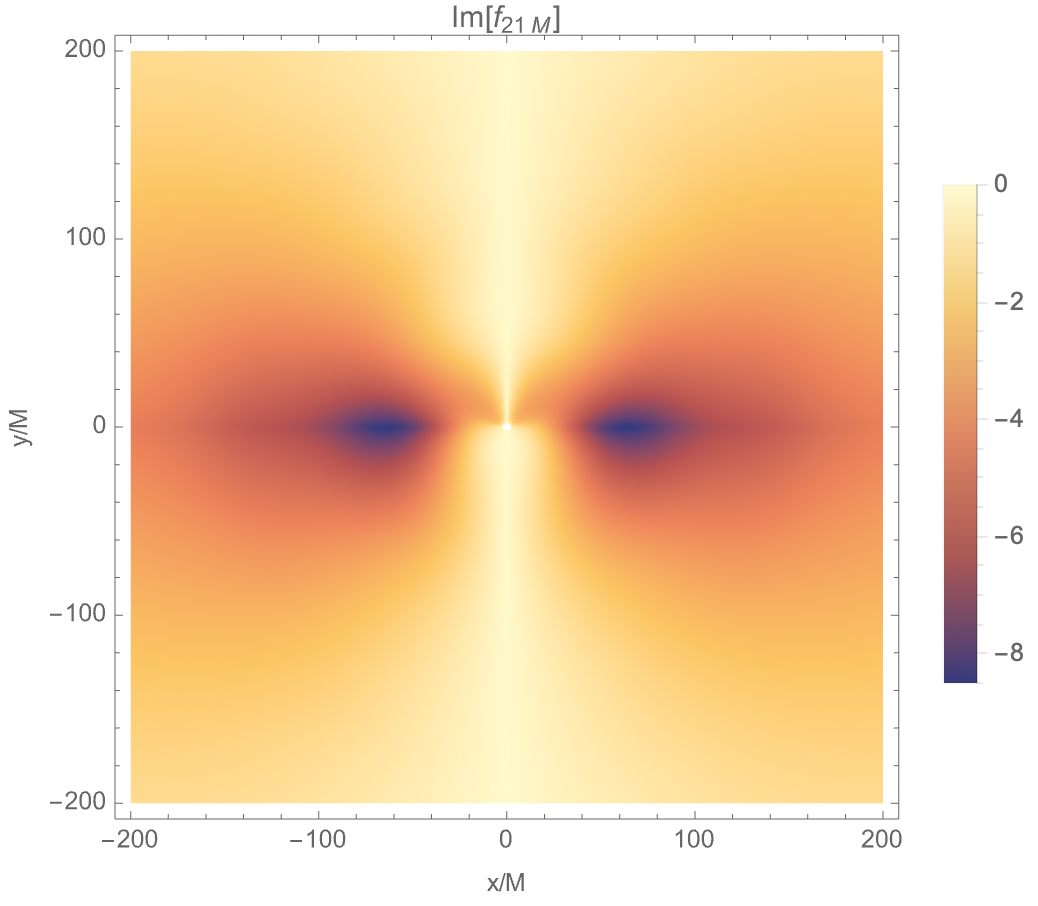}
\caption{The real and imaginal parts of $f_{01M}\left(r_0,\theta\right)$, $f_{11M}\left(r_0,\theta\right)$, $f_{21M}\left(r_0,\theta\right)$ induced from $\phi_1$ are plotted with $a_0=0.5$ and $\alpha=0.1$ in the scale of $200M$, where the superradiant condition is satisfied $\alpha<\frac{a_0}{2r_{\rm h0}}$.}
\label{Figures_f012}
\end{figure}

From the analysis of the equations, we found that the induced NP scalars from the two parts $\phi_1$ and $\phi_2$ of the axion cloud have following relations (here we remove the coefficient $M$)
\begin{eqnarray}
    &&{\rm Re}\left[f_{01}\left(r,\theta\right)\right]={\rm Re}\left[f_{02}\left(r,\pi-\theta\right)\right]~,\quad {\rm Im}\left[f_{01}\left(r,\theta\right)\right]=-{\rm Im}\left[f_{02}\left(r,\pi-\theta\right)\right]~, \label{relationf0}\\
    &&{\rm Re}\left[f_{11}\left(r,\theta\right)\right]=-{\rm Re}\left[f_{12}\left(r,\pi-\theta\right)\right]~,\quad {\rm Im}\left[f_{11}\left(r,\theta\right)\right]={\rm Im}\left[f_{12}\left(r,\pi-\theta\right)\right]~, \label{relationf1}\\
    &&{\rm Re}\left[f_{21}\left(r,\theta\right)\right]={\rm Re}\left[f_{22}\left(r,\pi-\theta\right)\right]~,\quad {\rm Im}\left[f_{21}\left(r,\theta\right)\right]=-{\rm Im}\left[f_{22}\left(r,\pi-\theta\right)\right]~, \label{relationf2}
\end{eqnarray}
which means that with the above transformations, we can start from one set of the equations to the other set. These symmetries from analysis are consistent with the numerical results, thus here we only exhibit the distribution for those from $\phi_1$ in FIG. \ref{Figures_f012}, where the axes $x=r_0 \sin{\theta}$ and $y=r_0 \cos{\theta}$ are defined to describe the cross section.

From FIG. \ref{Figures_f012} we can see that, the ingoing radiation $f_{01M}$ is slightly larger than the outgoing radiation $f_{21M}$, and the residential part $f_{11M}$ presents a ripple-like pattern. Usually the residential $\varphi_1$ part is neglected, since its equation is not separable even without any source and it is not related to ingoing or outgoing part. But here, it is shown that the imaginary part of $f_{11M}$ gets much larger amplitude than the other parts, indicating that the residential part is the primary constituent of the EM photon cloud we are concerned with. In fact, both $\varphi_0$ and $\varphi_2$ contain ingoing and outgoing parts, people assign ingoing/outgoing radiation to $\varphi_0$/$\varphi_2$ due to which one is dominant at spatial infinity. 

Subsequently, we can construct the whole EM tensor induced from $\phi_1$ and $\phi_2$ as
\begin{equation}
    F_{\mu\nu}^{(1)}=\cos{\left(m \varphi -\omega t\right)}F_{\mu\nu}^{\rm cos}\left(r,\theta\right)+\sin{\left(m \varphi -\omega t\right)}F_{\mu\nu}^{\rm sin}\left(r,\theta\right)~, \label{inducedF}
\end{equation}
where
\begin{eqnarray}
    F_{\mu\nu}^{\rm cos}\left(r,\theta\right)&=&4~{\rm Re}\left[\left(f_{01} +f_{02} \right)\overline{m}_{[\mu}n_{\nu ]}+\left(f_{11} +f_{12} \right)\left(n_{[ \mu}l_{\nu ]}+m_{[ \mu}\overline{m}_{\nu ]}\right)+\left(f_{21} +f_{22} \right)l_{[\mu}m_{\nu ]}\right]~,\label{f11}\\
    F_{\mu\nu}^{\rm sin}\left(r,\theta\right)&=&4~{\rm Im}\left[\left(f_{02} -f_{01} \right)\overline{m}_{[\mu}n_{\nu ]}+\left(f_{12} -f_{11} \right)\left(n_{[ \mu}l_{\nu ]}+m_{[ \mu}\overline{m}_{\nu ]}\right)+\left(f_{22} -f_{21} \right)l_{[\mu}m_{\nu ]}\right]~.\label{f22}
\end{eqnarray}

So far we have considered an EM photon cloud sourced by an axion cloud surrounding a Kerr black hole which is described by (\ref{inducedF}). This is one of the main results of our work. Moreover, it means that for an arbitrary current $J^a=\Sigma e^{\pm i\omega t \mp im\varphi}S\left(r,\theta\right)$ as the source we will always get a photon cloud evolving as (\ref{inducedF}). In fact, even photons without any couplings can acquire an effective mass in a plasma, allowing superradiance to occur for some primordial black holes~\cite{Pani:2013hpa,Fukuda:2019ewf}.

\begin{figure}[h]
\centering
  \includegraphics[width=0.3\textwidth]{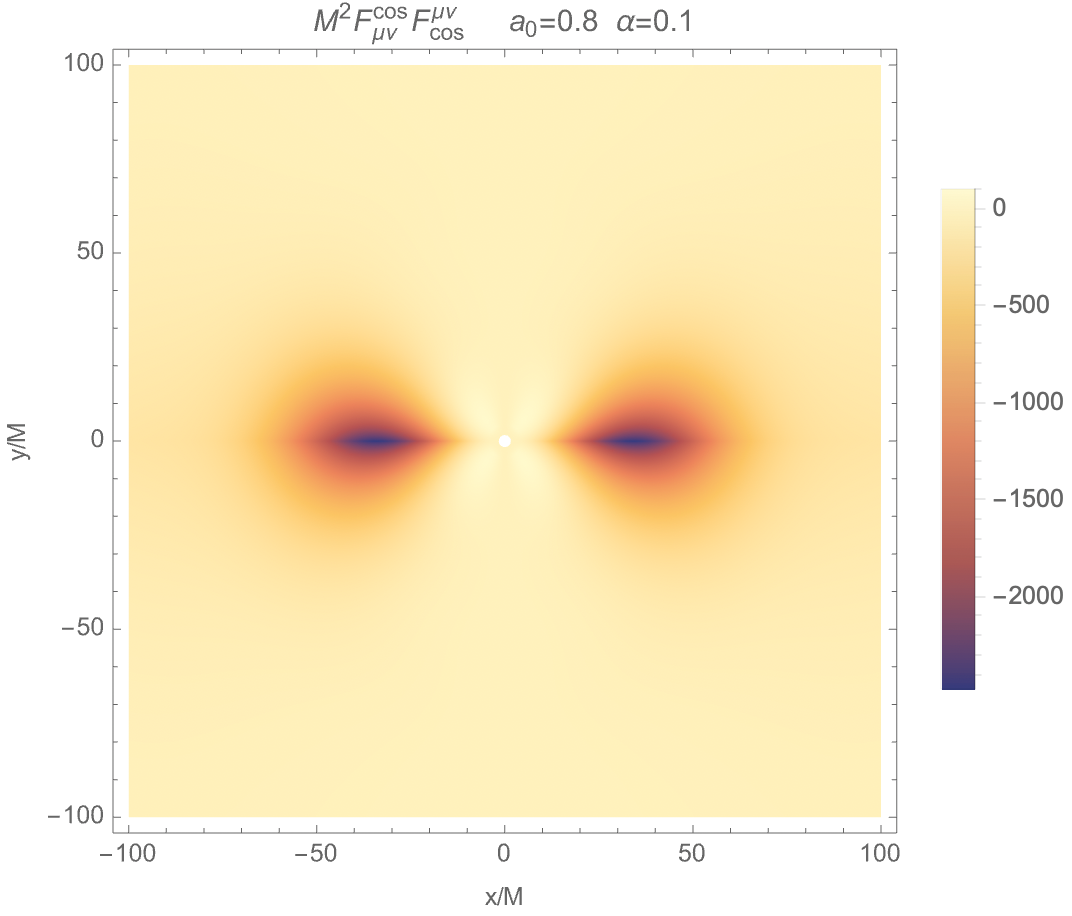}  
  \includegraphics[width=0.3\textwidth]{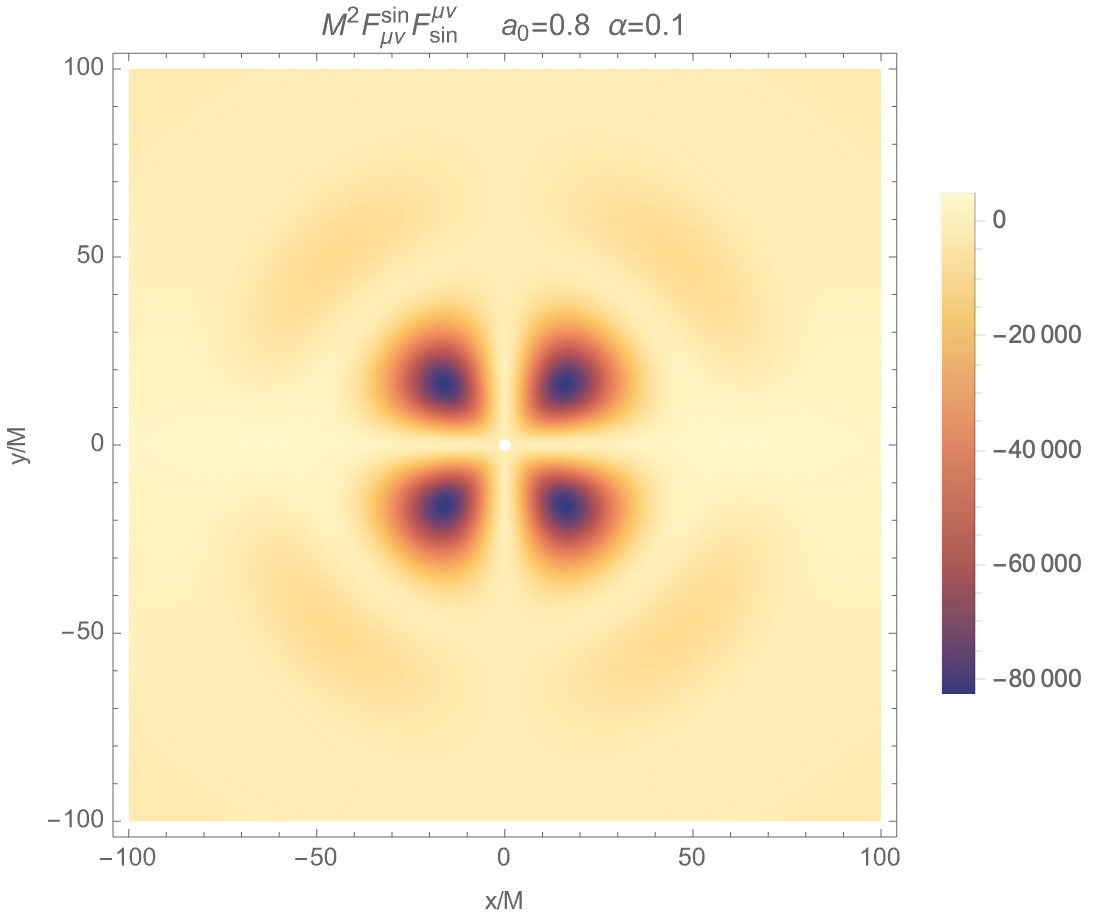} 
   \includegraphics[width=0.3\textwidth]{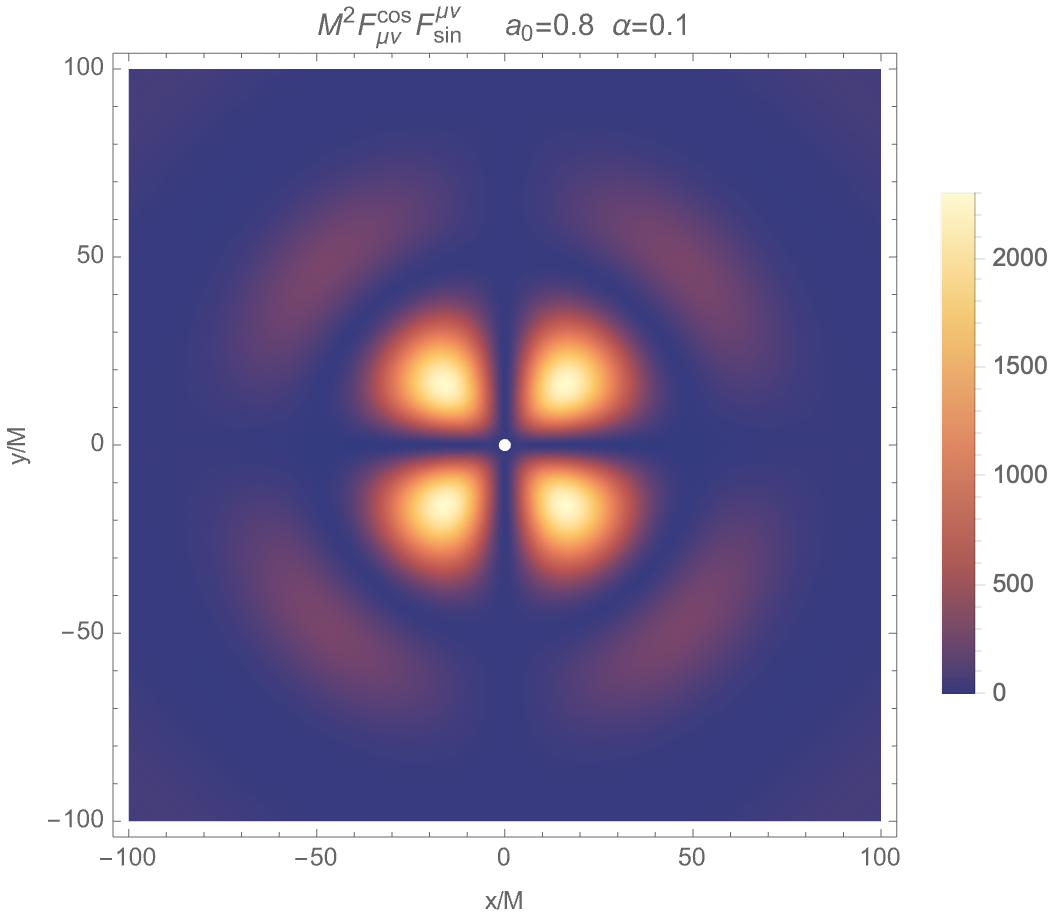} 
   \includegraphics[width=0.3\textwidth]{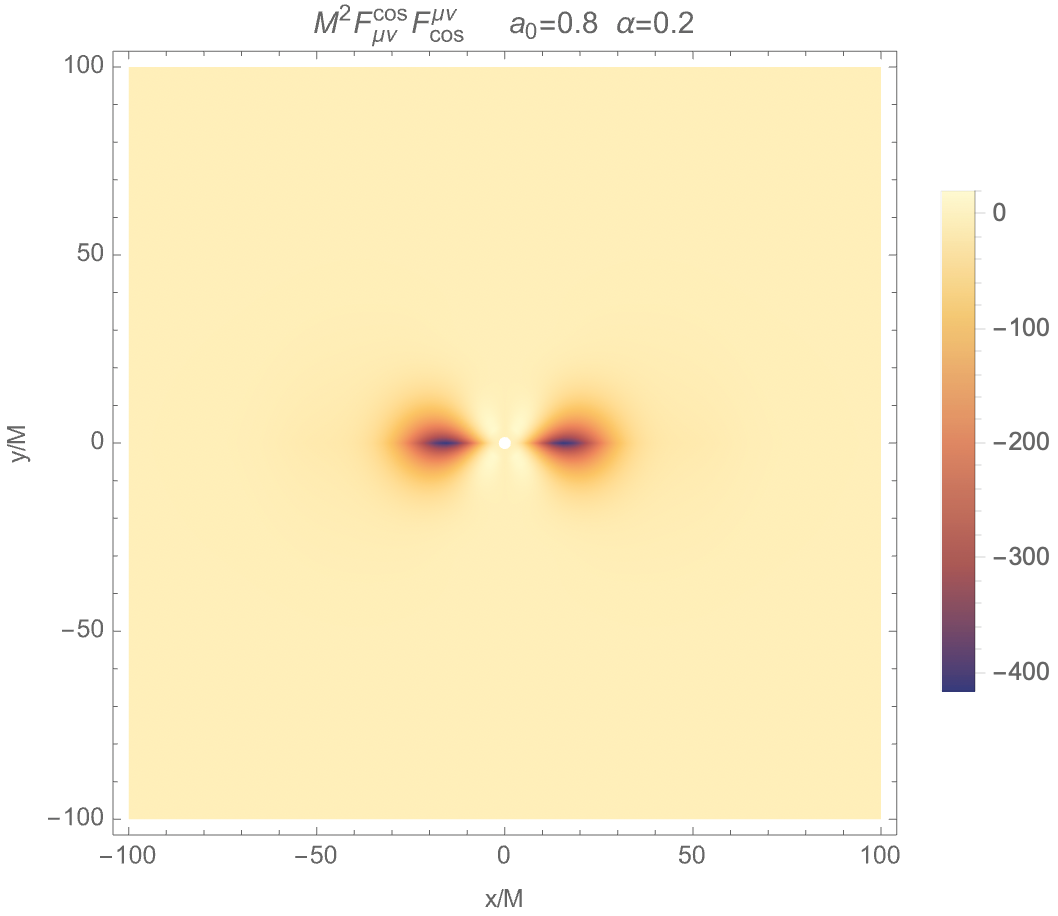} 
   \includegraphics[width=0.3\textwidth]{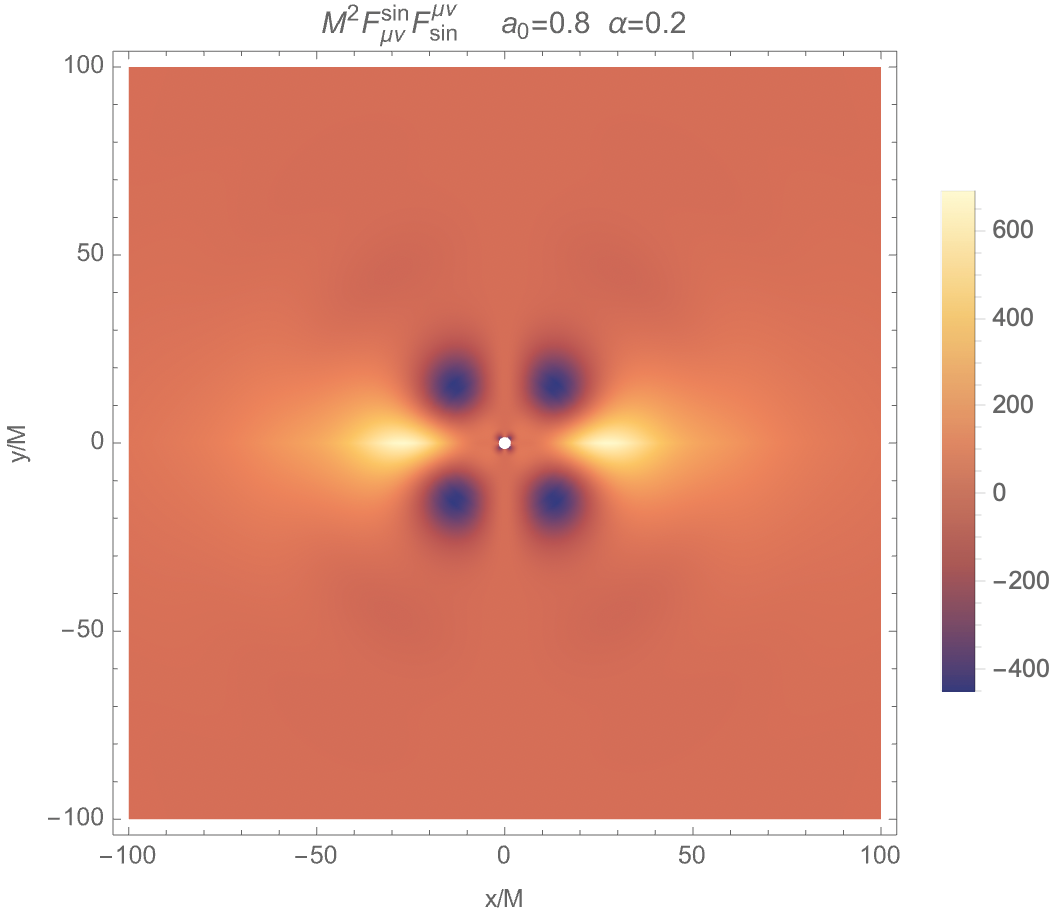} \includegraphics[width=0.3\textwidth]{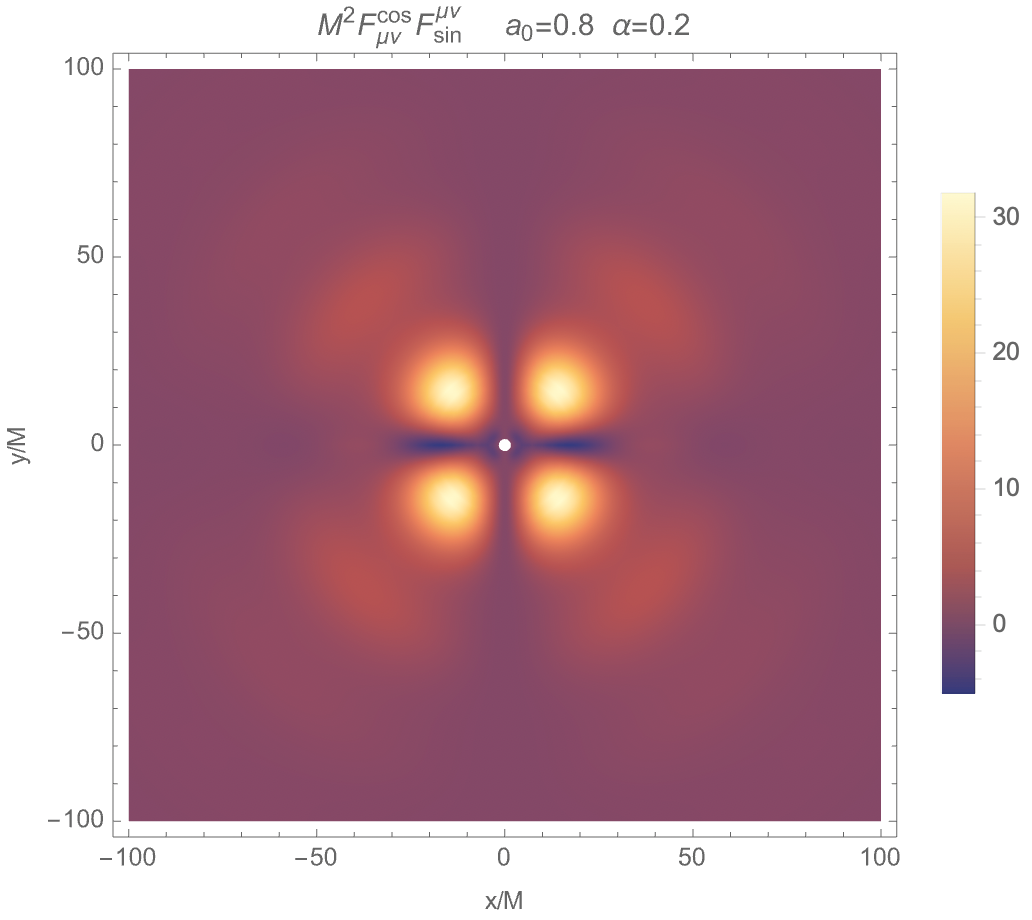}  
\caption{The components $M^2F_{\mu\nu}^{\rm cos}F^{\mu\nu}_{\rm cos},~M^2 F_{\mu\nu}^{\rm sin}F^{\mu\nu}_{\rm sin},~M^2F_{\mu\nu}^{\rm cos}F^{\mu\nu}_{\rm sin}$ of the contraction of the EM field strength tensor $M^2 F_{\mu\nu}^{(1)}F^{\mu\nu}_{(1)}$ are plotted with $a_0=0.8$ and $\alpha=0.1,0.2$.}
\label{Figures_FF}
\end{figure}

In~\cite{Spieksma:2023vwl}, a detailed study of the coupled axion-Maxwell system was performed on a black hole background, showing that, in the presence of plasmas, the EM instability remains controlled by the axion-photon coupling, even for relatively dense plasmas. In our case, the frequency of the induced photon cloud is twice the EM frequency in their work, making it less susceptible to plasma effects. Moreover, as shown in FIG. \ref{Figures_FF}, the induced EM scalar $M^2 F_{\mu\nu}^{(1)}F^{\mu\nu}_{(1)}$ is concentrated in regions away from both the equatorial plane and the polar axis, which renders it less affected by the accretion disk and relativistic jets. 

Importantly, the photon component of the cloud does not correspond to an on-shell propagating photon mode, but instead constitutes part of a mixed axion–photon propagation eigenstate undergoing superradiant amplification. Consequently, the existence of a photon component is not excluded by plasma screening. 

Furthermore, we emphasize that although the induced photon cloud is initially bound, it corresponds to unstable bound configurations that can subsequently escape to spatial infinity. Since photons experience no potential barrier at spatial infinity where the contribution from the axion cloud vanishes. From a particle interpretation, the photon component can be regarded as effectively free once it is released from the axion cloud into propagating EM modes.

We now turn to the evolution of the induced EM photon cloud. The parity violation term $F_{\mu\nu}^{(1)}{}^*F^{\mu\nu}_{(1)}$ and the contraction of the EM field strength tensor $F_{\mu\nu}^{(1)}F^{\mu\nu}_{(1)}$ can be obtained as ($m=1$)
\begin{eqnarray}
    F_{\mu\nu}^{(1)}{}^*F^{\mu\nu}_{(1)}&=&\cos^2{\left(m \varphi -\omega t\right)}F_{\mu\nu}^{\rm cos}{}^*F^{\mu\nu}_{\rm cos}+\sin^2{\left(m \varphi -\omega t\right)}F_{\mu\nu}^{\rm sin}{}^*F^{\mu\nu}_{\rm sin}+2\cos{\left(m \varphi -\omega t\right)}\sin{\left(m \varphi -\omega t\right)}F_{\mu\nu}^{\rm cos}{}^*F^{\mu\nu}_{\rm sin}~\\
    &=&\cos{\left(2m \varphi -2\omega t\right)}\left(\frac{1}{2}F_{\mu\nu}^{\rm cos}{}^*F^{\mu\nu}_{\rm cos}-\frac{1}{2}F_{\mu\nu}^{\rm sin}{}^*F^{\mu\nu}_{\rm sin}\right)+\sin{\left(2m \varphi -2\omega t\right)}F_{\mu\nu}^{\rm cos}{}^*F^{\mu\nu}_{\rm sin}+\frac{1}{2}F_{\mu\nu}^{\rm cos}{}^*F^{\mu\nu}_{\rm cos}+\frac{1}{2}F_{\mu\nu}^{\rm sin}{}^*F^{\mu\nu}_{\rm sin}~, \notag \\
    F_{\mu\nu}^{(1)}F^{\mu\nu}_{(1)}&=&\cos^2{\left(m \varphi -\omega t\right)}F_{\mu\nu}^{\rm cos}F^{\mu\nu}_{\rm cos}+\sin^2{\left(m \varphi -\omega t\right)}F_{\mu\nu}^{\rm sin}F^{\mu\nu}_{\rm sin}+2\cos{\left(m \varphi -\omega t\right)}\sin{\left(m \varphi -\omega t\right)}F_{\mu\nu}^{\rm cos}F^{\mu\nu}_{\rm sin}~\\
    &=&\cos{\left(2m \varphi -2\omega t\right)}\left(\frac{1}{2}F_{\mu\nu}^{\rm cos}F^{\mu\nu}_{\rm cos}-\frac{1}{2}F_{\mu\nu}^{\rm sin}F^{\mu\nu}_{\rm sin}\right)+\sin{\left(2m \varphi -2\omega t\right)}F_{\mu\nu}^{\rm cos}F^{\mu\nu}_{\rm sin}+\frac{1}{2}F_{\mu\nu}^{\rm cos}F^{\mu\nu}_{\rm cos}+\frac{1}{2}F_{\mu\nu}^{\rm sin}F^{\mu\nu}_{\rm sin}~, \notag
\end{eqnarray}
which shows that except for the evolving parts with a double frequency $2\omega$, there are also residential parts that are time-independent. The $r,\theta$-dependent components $F_{\mu\nu}^{\rm cos}{}^*F^{\mu\nu}_{\rm cos}$, $F_{\mu\nu}^{\rm sin}{}^*F^{\mu\nu}_{\rm sin}$, $F_{\mu\nu}^{\rm cos}{}^*F^{\mu\nu}_{\rm sin}$, $F_{\mu\nu}^{\rm cos}F^{\mu\nu}_{\rm cos}$, $F_{\mu\nu}^{\rm sin}F^{\mu\nu}_{\rm sin}$ and $F_{\mu\nu}^{\rm cos}F^{\mu\nu}_{\rm sin}$ are plotted in FIG. \ref{Figures_FFD}-\ref{Figures_FF}. Their expressions can be found in Appendix \ref{FF}.

\begin{figure}[h]
\centering
 \includegraphics[width=0.3\textwidth]{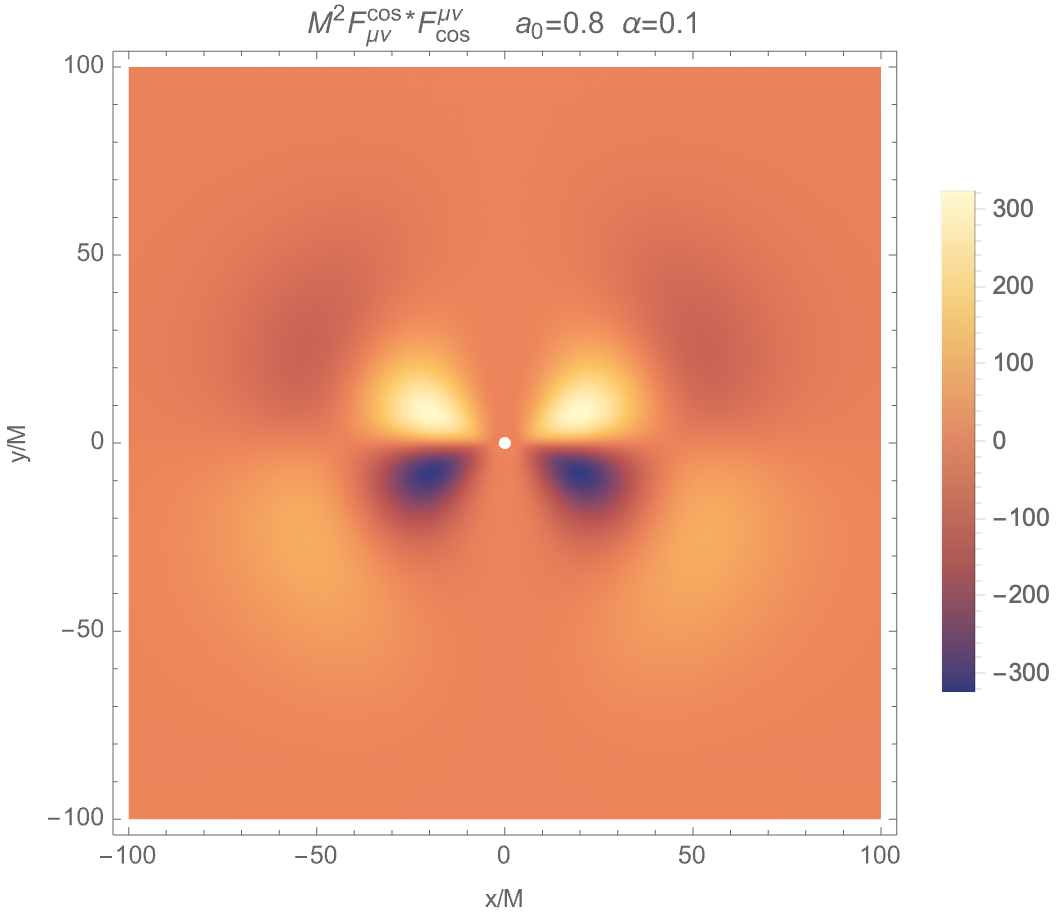}  
 \includegraphics[width=0.3\textwidth]{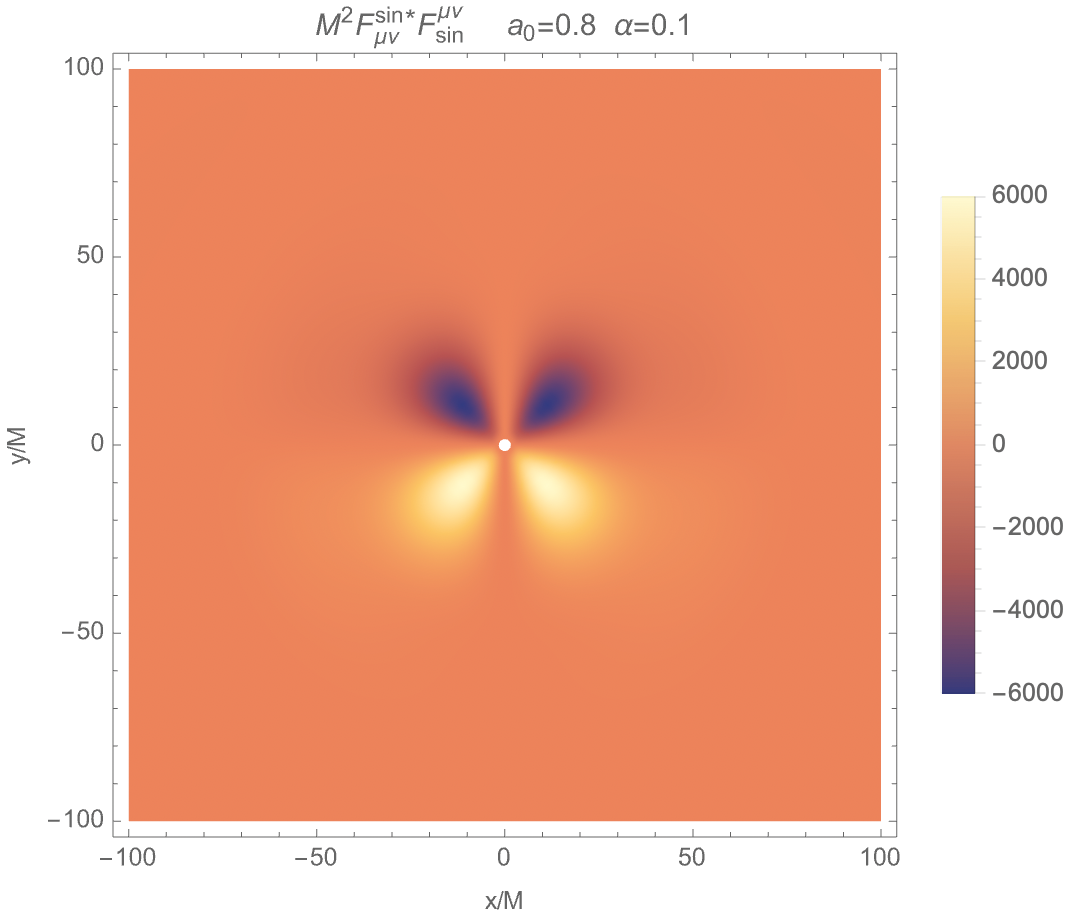} 
 \includegraphics[width=0.3\textwidth]{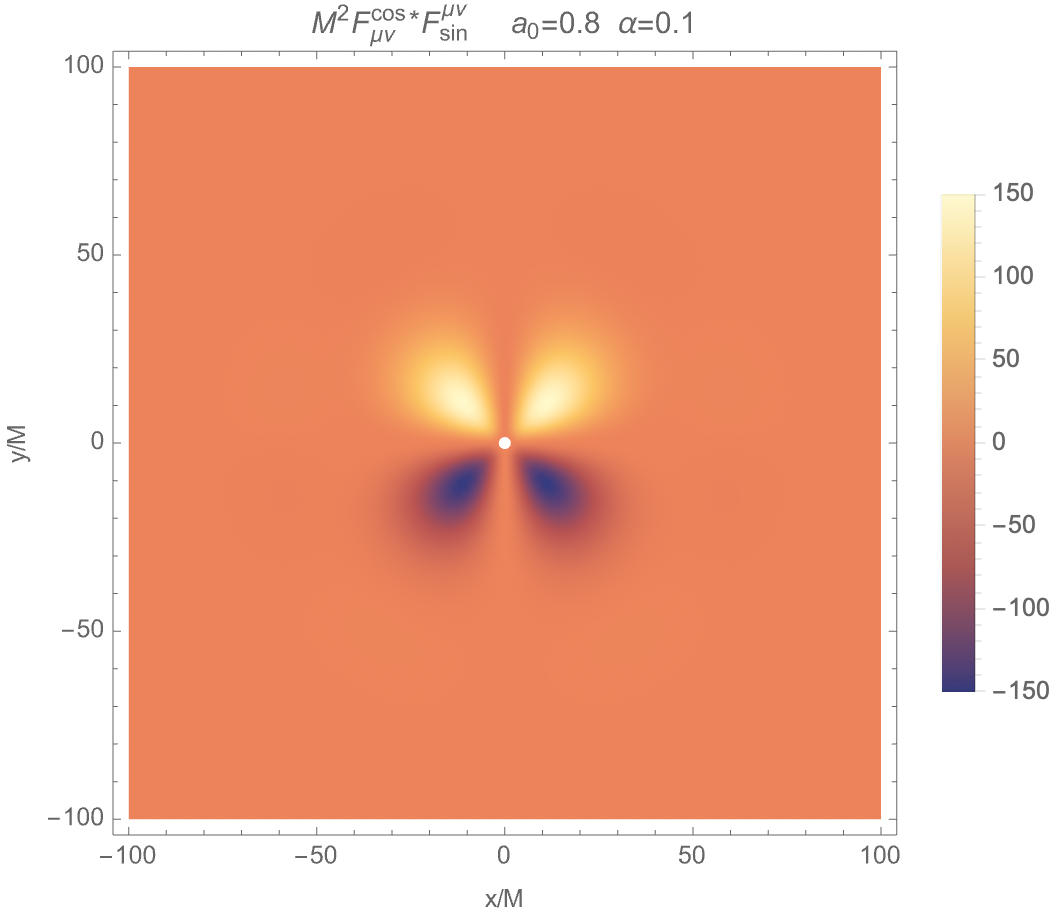}  \includegraphics[width=0.3\textwidth]{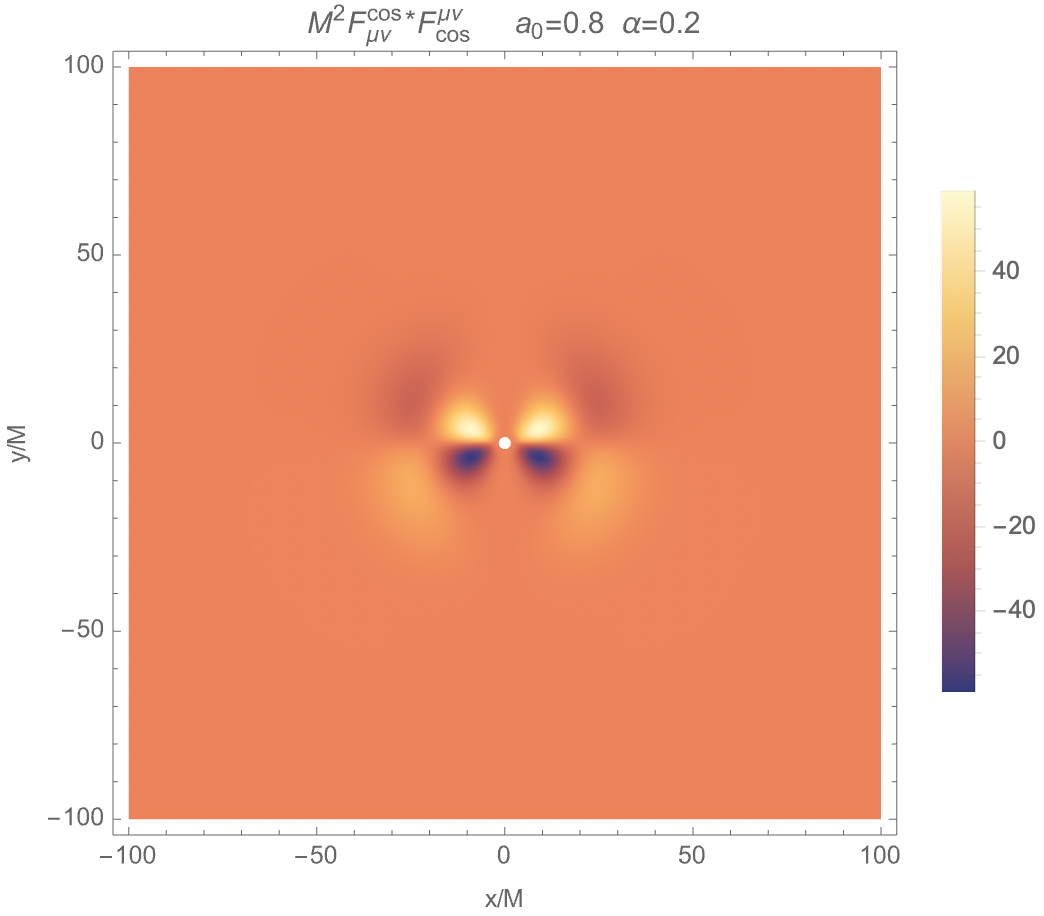}  
 \includegraphics[width=0.3\textwidth]{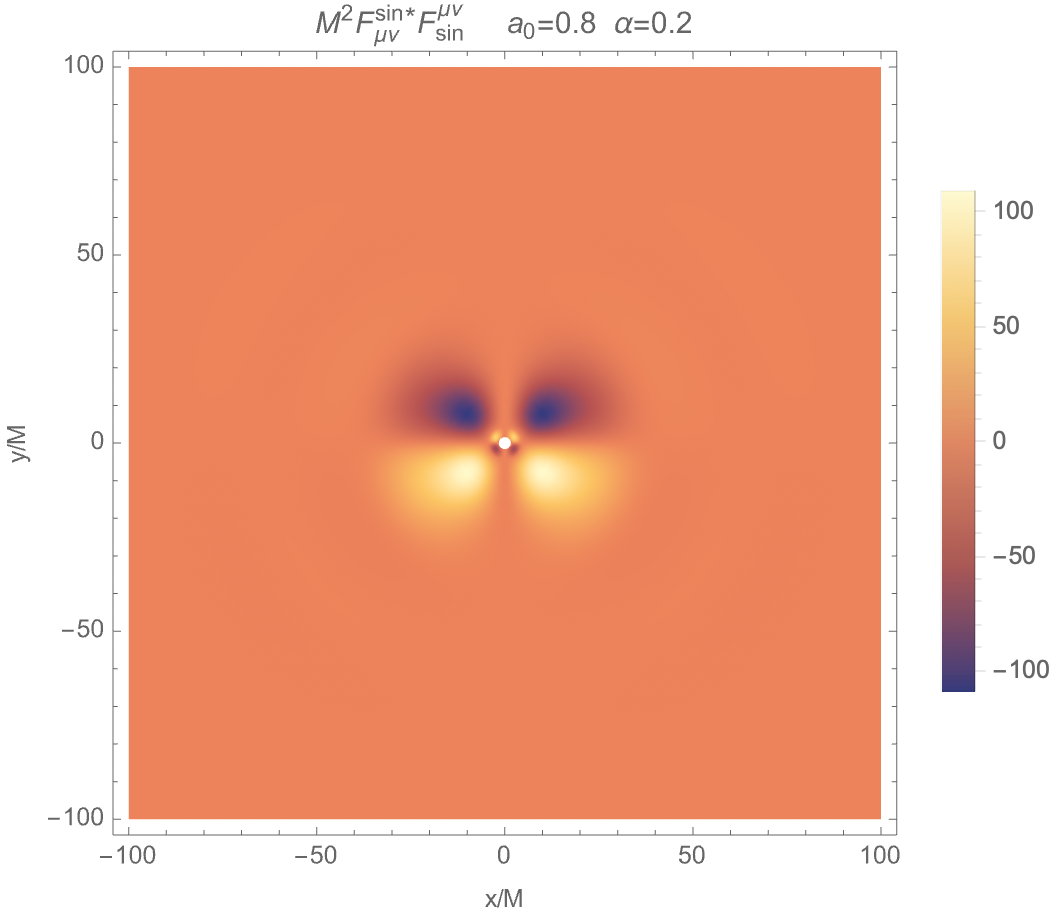}  
 \includegraphics[width=0.3\textwidth]{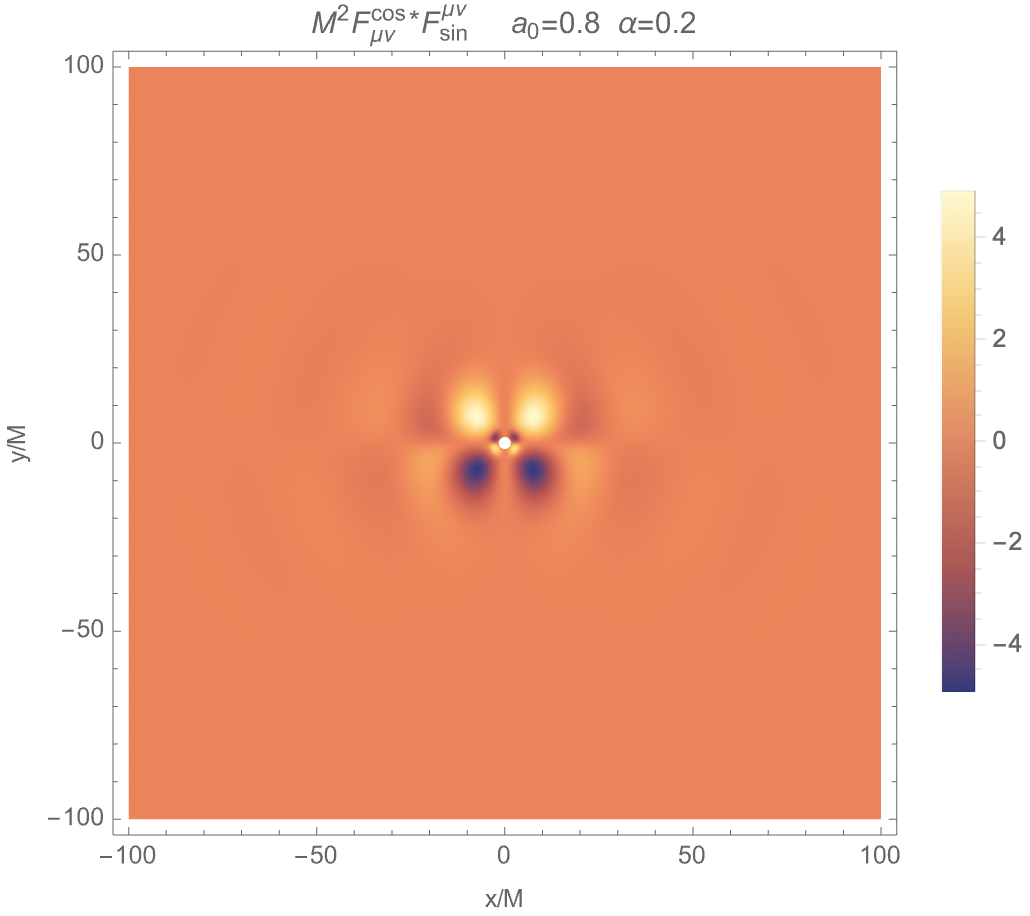}  
\caption{The components $M^2 F_{\mu\nu}^{\rm cos}{}^*F^{\mu\nu}_{\rm cos},~M^2 F_{\mu\nu}^{\rm sin}{}^*F^{\mu\nu}_{\rm sin},~M^2 F_{\mu\nu}^{\rm cos}{}^*F^{\mu\nu}_{\rm sin}$ of the parity violation term $M^2 F_{\mu\nu}^{(1)}{}^*F^{\mu\nu}_{(1)}$ are plotted with $a_0=0.8$ and $\alpha=0.1,0.2$.}
\label{Figures_FFD}
\end{figure}

We study the effects of the two parameters $a_0,~\alpha$ and find that the spin parameter $a_0$ only has slight impacts on both the amplitude and the structure of the distribution for each components of $M^2 F_{\mu\nu}^{(1)}{}^*F^{\mu\nu}_{(1)}$ and $M^2 F_{\mu\nu}^{(1)}F^{\mu\nu}_{(1)}$, while these quantities are sensitive to the fine structure constant $\alpha$. Therefore, in FIG. \ref{Figures_FFD}-\ref{Figures_FF} we plot the figures for the components of $M^2 F_{\mu\nu}^{(1)}{}^*F^{\mu\nu}_{(1)}$ and $M^2 F_{\mu\nu}^{(1)}F^{\mu\nu}_{(1)}$ with fixed $a_0=0.8$ and different values of $\alpha$. From the figures we can see that, for all the components, the increase of $\alpha$ leads to a significant decrease in amplitude and a slight concentration on the distribution, which indicates that for fixed black hole mass $M$ the smaller axion mass $\mu$ (larger reduced Compton wavelength) makes the interaction with photons more efficient and stronger. This is consistent with the idea that very light (or massless) particles may have a near-degeneracy with the photons, leading to a mixing phenomenon in which a coherent superposition of the two arises~\cite{Raffelt:1987im}. Also, the slight concentration on the distribution should result from the concentration of the axion cloud with larger $\alpha$, where larger axion mass $\mu$ (smaller reduced Compton wavelength) makes the axion cloud concentrate. In general, the induced EM scalars $M^2 F_{\mu\nu}^{(1)}{}^*F^{\mu\nu}_{(1)}$ and $M^2 F_{\mu\nu}^{(1)}F^{\mu\nu}_{(1)}$ concentrate in the sine parts $M^2F_{\mu\nu}^{\rm sin}{}^*F^{\mu\nu}_{\rm sin}$ and $M^2F_{\mu\nu}^{\rm sin}F^{\mu\nu}_{\rm sin}$.

It is essential to verify the validity range of the perturbative framework with respect to the EM field, ensuring that $F_{\mu\nu}^{(0)}\gg \varepsilon F_{\mu\nu}^{(1)}$. Using $F_{\mu\nu}^{(0)}\sim B_0$, this requirement leads to
\begin{equation}
    \alpha^2 \phi_0 k_a MF_{\mu\nu}^{(1)}\ll 1~. \label{condition}
\end{equation}
For the case with $a_0=0.8$ and $\alpha=0.1$, the maximum value of $M^2F_{\mu\nu}^{(1)}F^{\mu\nu}_{(1)}$ reaches approximately $8\times 10^4$, which gives $\phi_0 k_a \ll  0.35$. While for the case with  $a_0=0.8$ and $\alpha=0.2$, the constraint is significantly relaxed as $\phi_0 k_a \ll 1.25$. Note that the quantity $\phi_0 k_a$ is dimensionless in both geometrized and natural units. Remarkably, the resulting constraint is independent of the background field strength $B_0$.

\section{The induced electric and magnetic fields}
\label{The induced electric and magnetic fields}

In this Section we will study the induced electric field and magnetic field respectively. Considering the $(3+1)$-decomposition of the spacetime~\cite{Alcubierr:2008}
\begin{equation}
    ds^2=-\alpha^2dt^2+\gamma_{ij}\left(dx^i+\beta^i dt\right)\left(dx^j+\beta^j dt\right)~,
\end{equation}
where $\beta^i$ is shift vector, $\gamma_{ij}$ is the spatial metric and the lapse function $\alpha=\sqrt{\frac{\Delta_r \Sigma}{\Delta_r \Sigma+2Mr\left(a^2+r^2\right)}}$ in Kerr metric (\ref{metric}). We can define the electric field and magnetic field~\cite{Alcubierre:2009ij}
\begin{equation}
    E^a :=n_b F^{ab}~, \quad B^a:=n_b{}^*F^{ab}~,
\end{equation}
with respect to an Eulerian observer $n^a=\left(1/\alpha,-\beta^i/\alpha\right)$ whose worldline is orthogonal to the spacelike hypersurface. Therefore we have 
\begin{eqnarray}
    E^i_{(1)}=\cos{\left(m \varphi -\omega t\right)}E^i_{\rm cos}\left(r,\theta\right)+\sin{\left(m \varphi -\omega t\right)}E^i_{\rm sin}\left(r,\theta\right)~, \\
    B^i_{(1)}=\cos{\left(m \varphi -\omega t\right)}B^i_{\rm cos}\left(r,\theta\right)+\sin{\left(m \varphi -\omega t\right)}B^i_{\rm sin}\left(r,\theta\right)~,
\end{eqnarray}
where
\begin{eqnarray}
    &&E^i_{\rm cos}=-\alpha F^{i0}_{\rm cos}~,\quad E^i_{\rm sin}=-\alpha F^{i0}_{\rm sin}~,\\
    &&B^i_{\rm cos}=-\frac{1}{2}\alpha \epsilon^{i0jk}F_{jk}^{\rm cos}~, \quad B^i_{\rm sin}=-\frac{1}{2}\alpha \epsilon^{i0jk}F_{jk}^{\rm sin}~.
\end{eqnarray}

\begin{figure}[h]
\centering
  \includegraphics[width=0.3\textwidth]{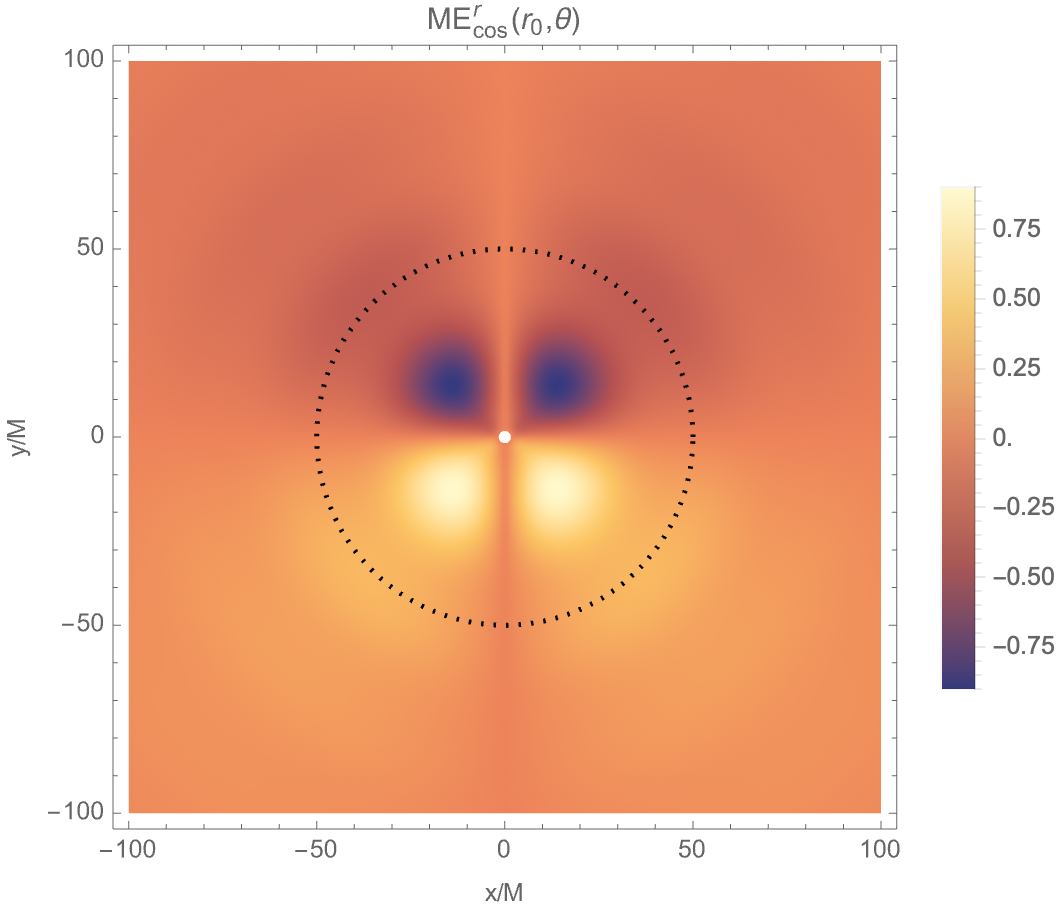}
 \includegraphics[width=0.3\textwidth]{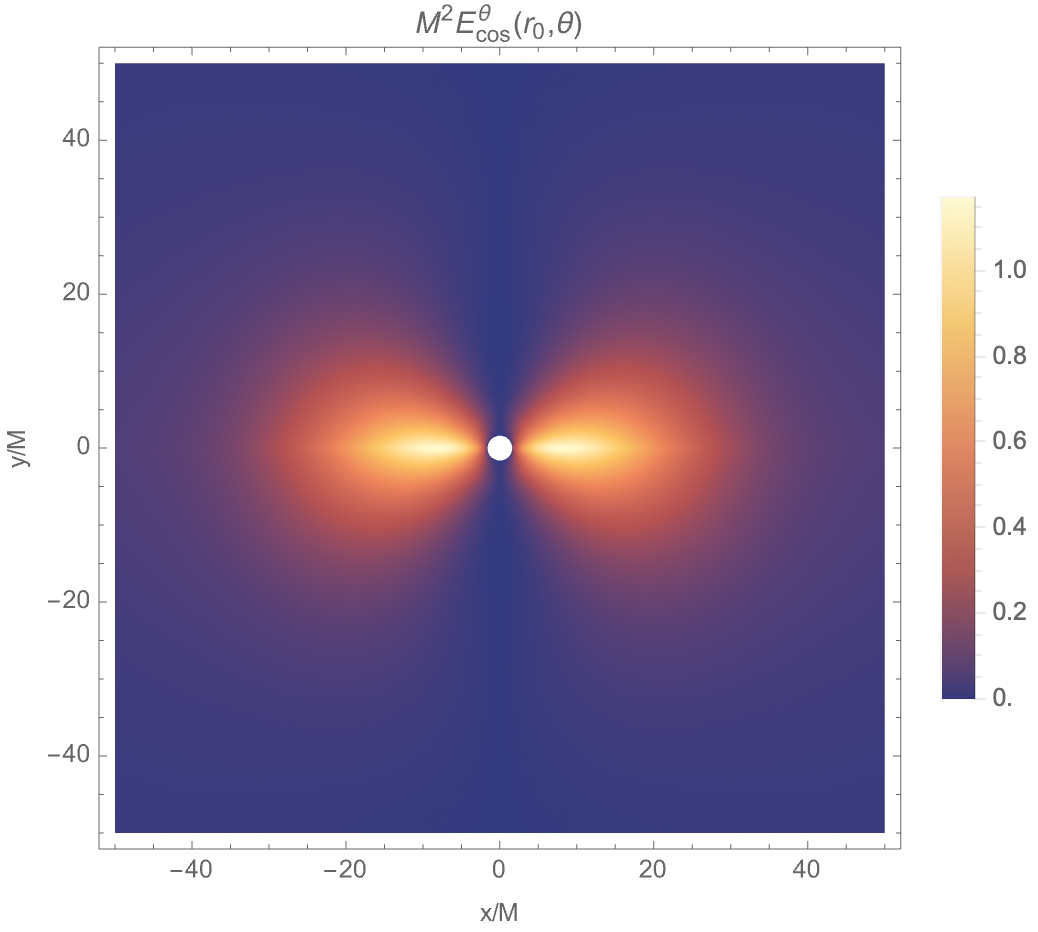}
  \includegraphics[width=0.3\textwidth]{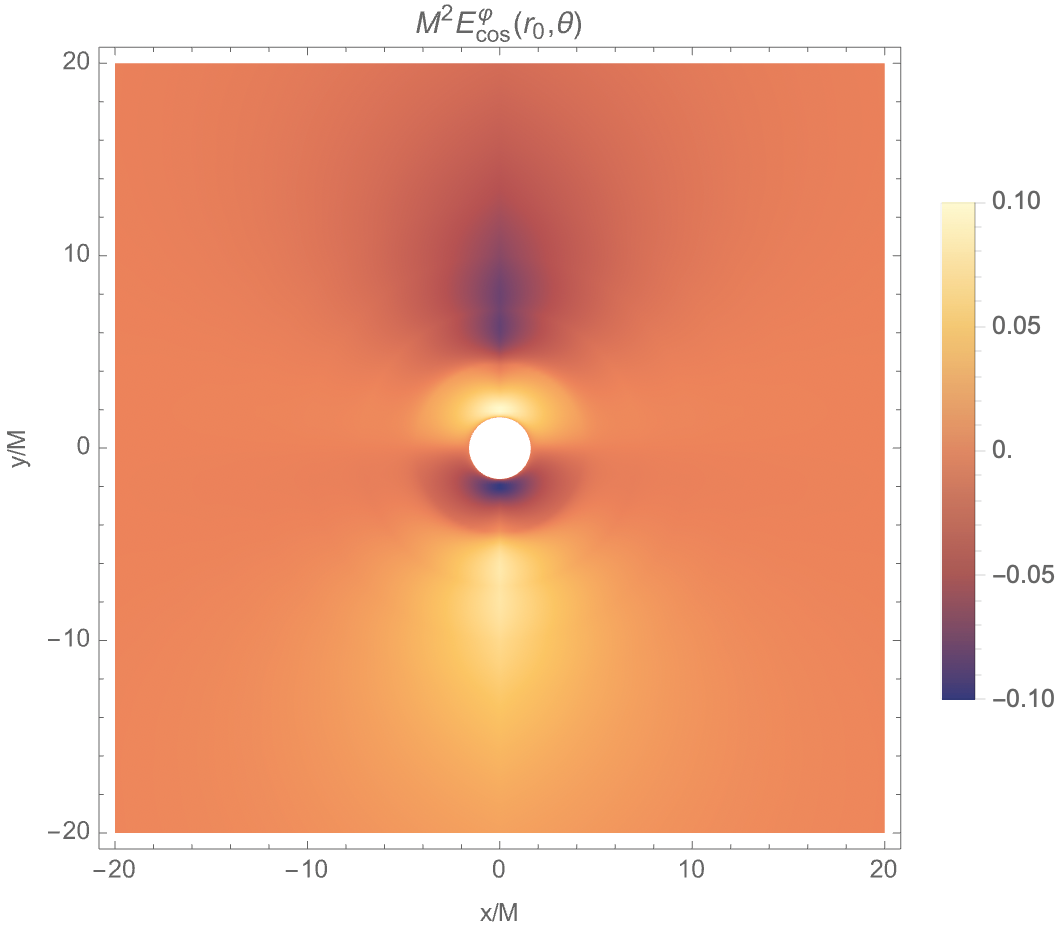}
 \includegraphics[width=0.3\textwidth]{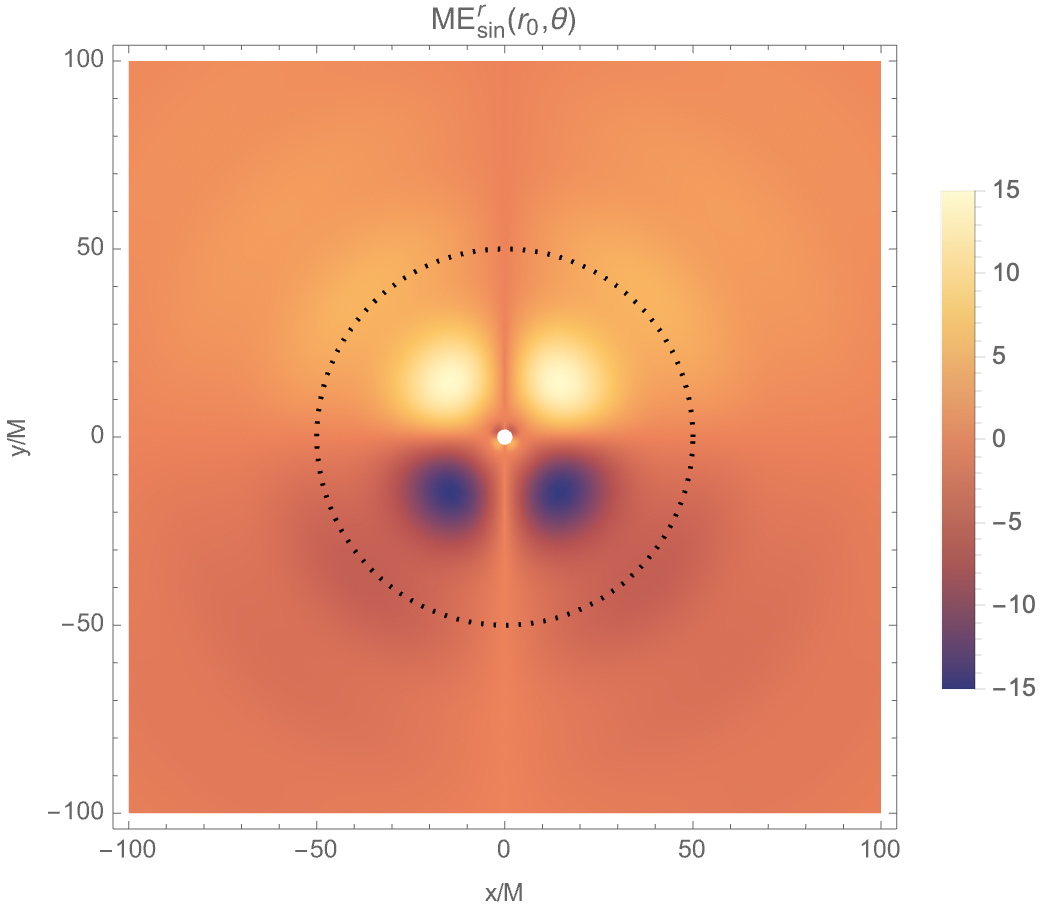}
 \includegraphics[width=0.3\textwidth]{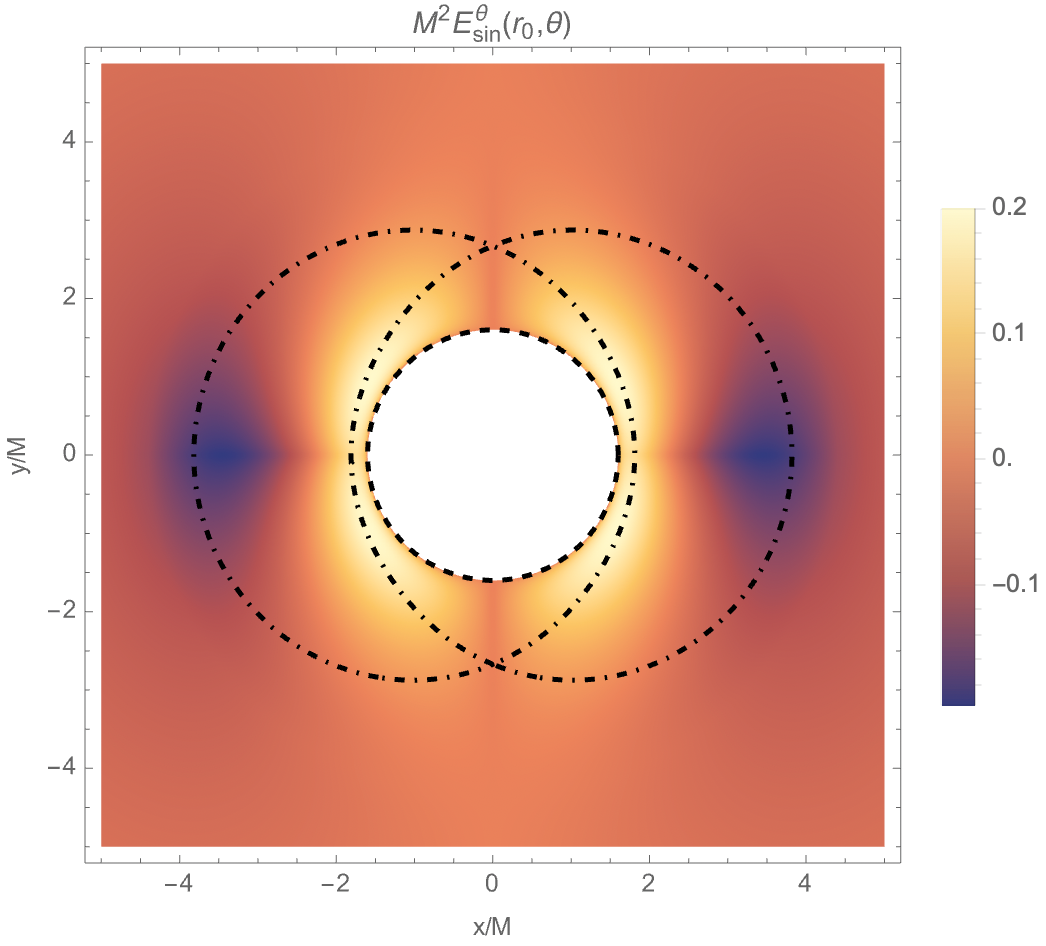}
  \includegraphics[width=0.3\textwidth]{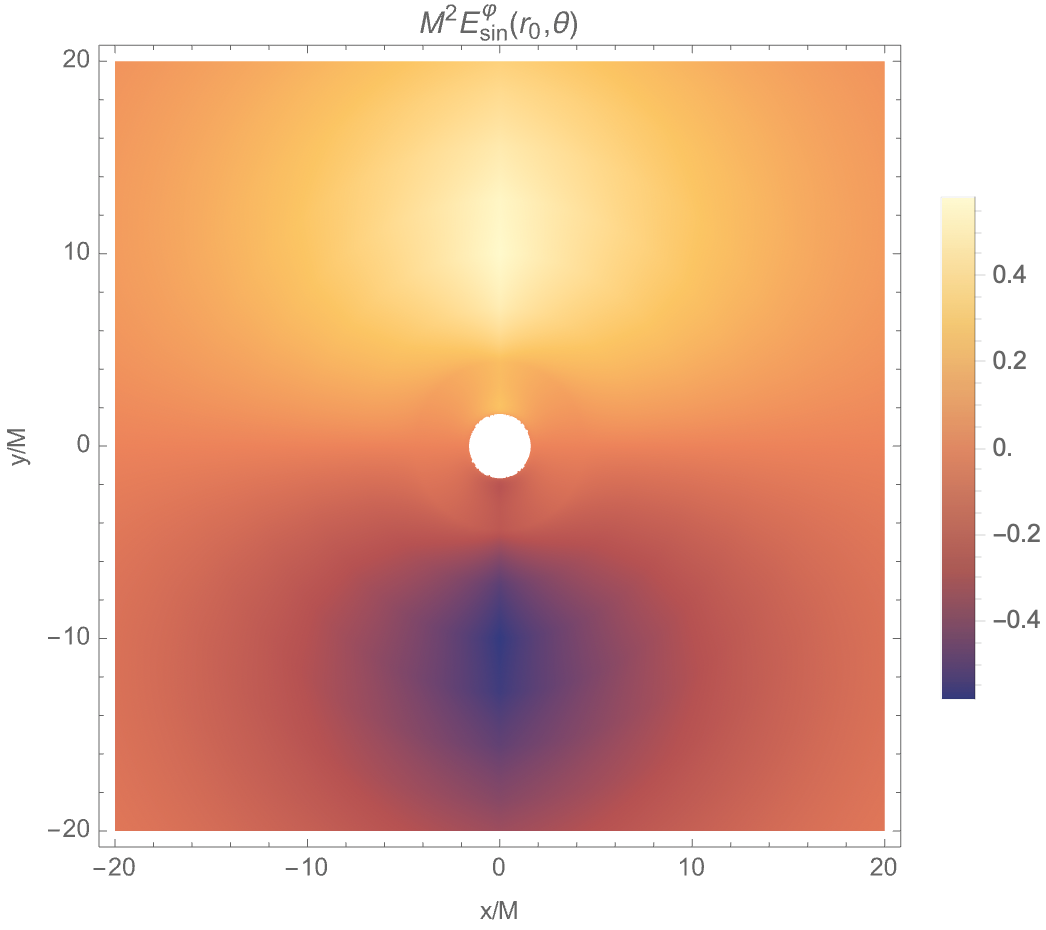}
\caption{The dimensionless quantities $M E^r_{(1)}$, $M^2 E^\theta_{(1)}$ and $M^2 E^\varphi_{(1)}$ of the induced electric field components are plotted as cosine part and sine part respectively. In all the figures we have set $a_0=0.8$ and $\alpha=0.2$. }
\label{Figures_E}
\end{figure}

\begin{figure}[h]
\centering
  \includegraphics[width=0.3\textwidth]{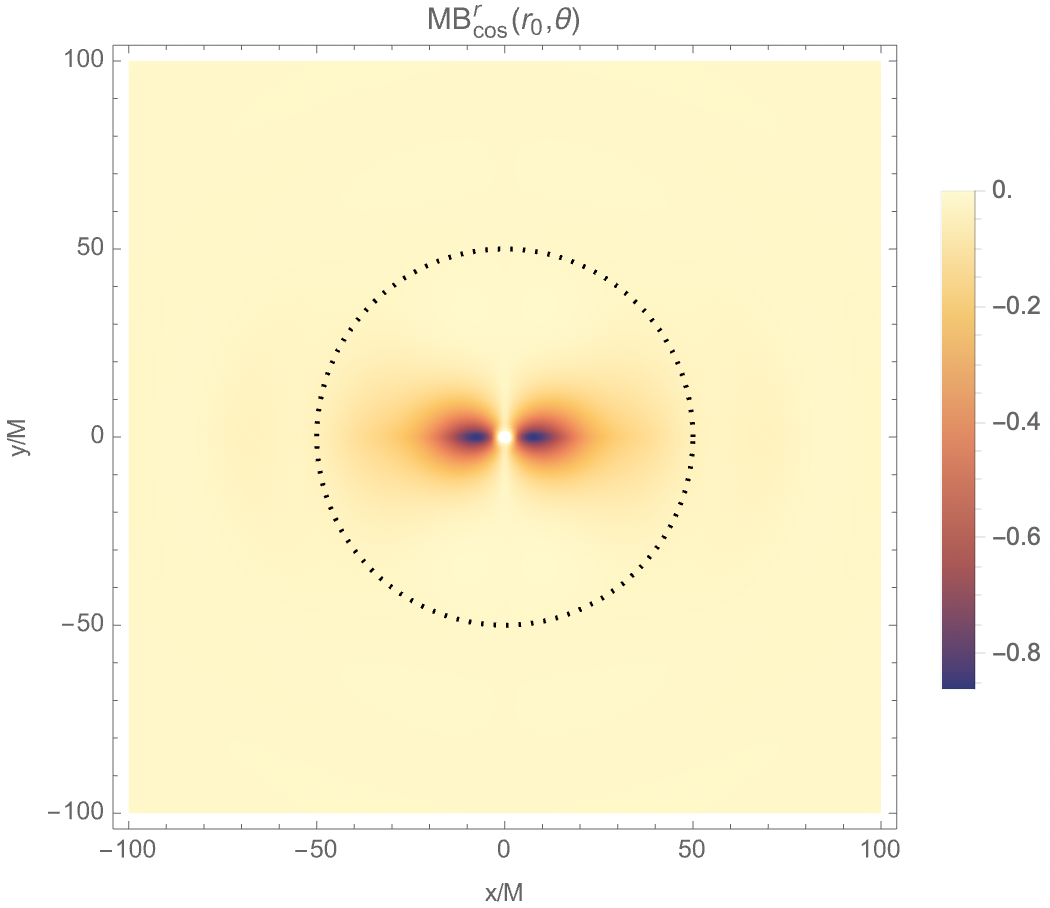}
  \includegraphics[width=0.3\textwidth]{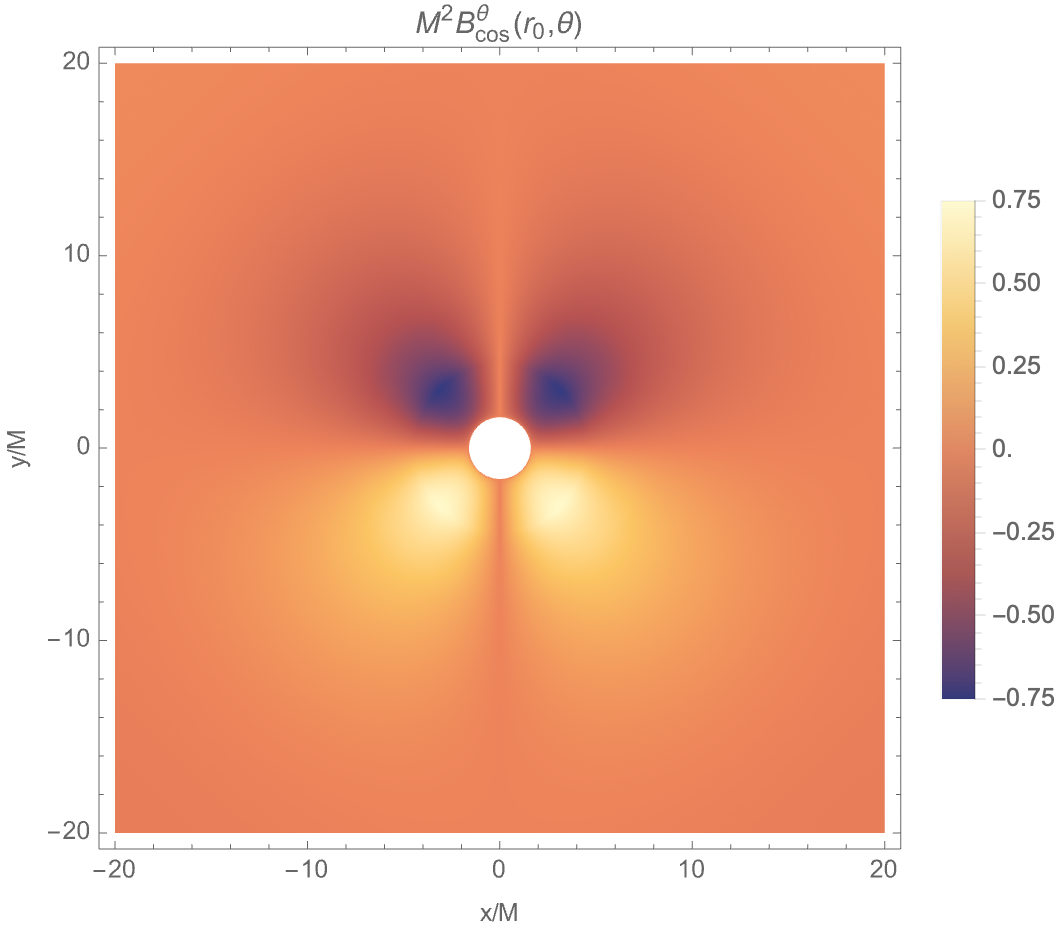}
  \includegraphics[width=0.3\textwidth]{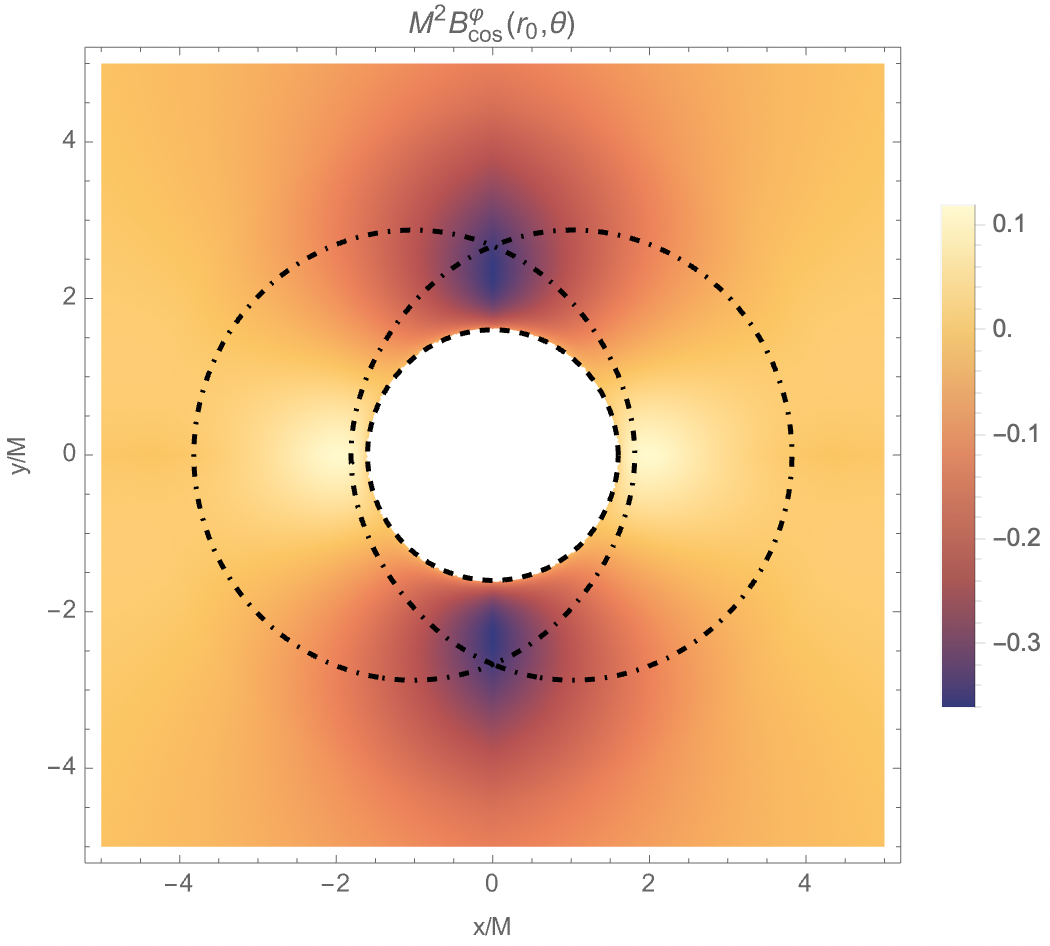}
  \includegraphics[width=0.3\textwidth]{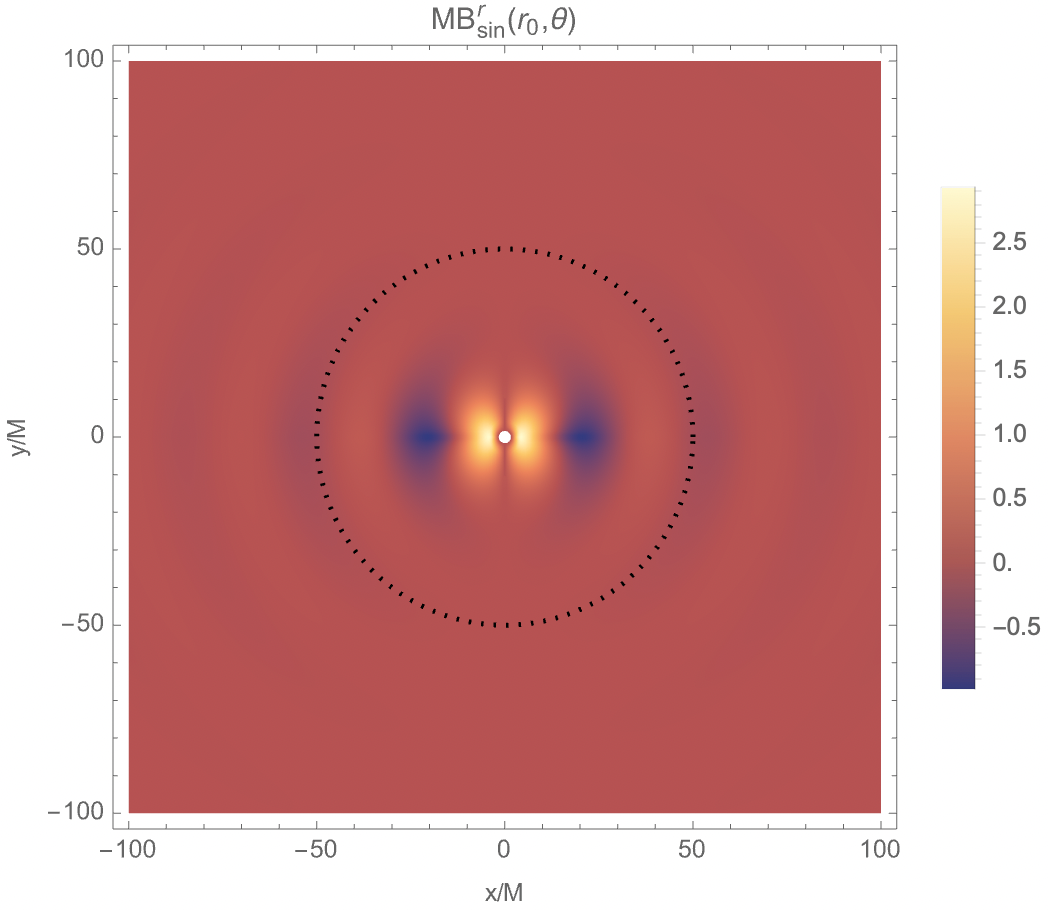}
  \includegraphics[width=0.3\textwidth]{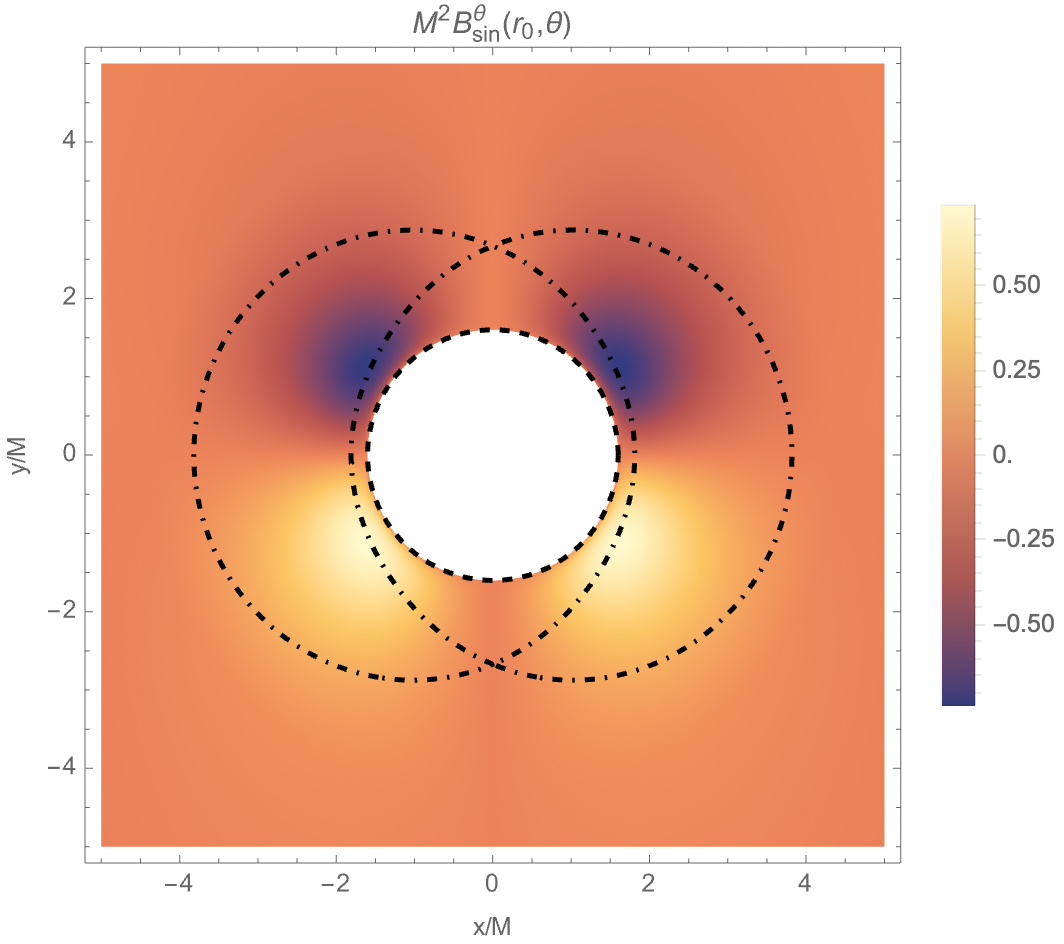}
  \includegraphics[width=0.3\textwidth]{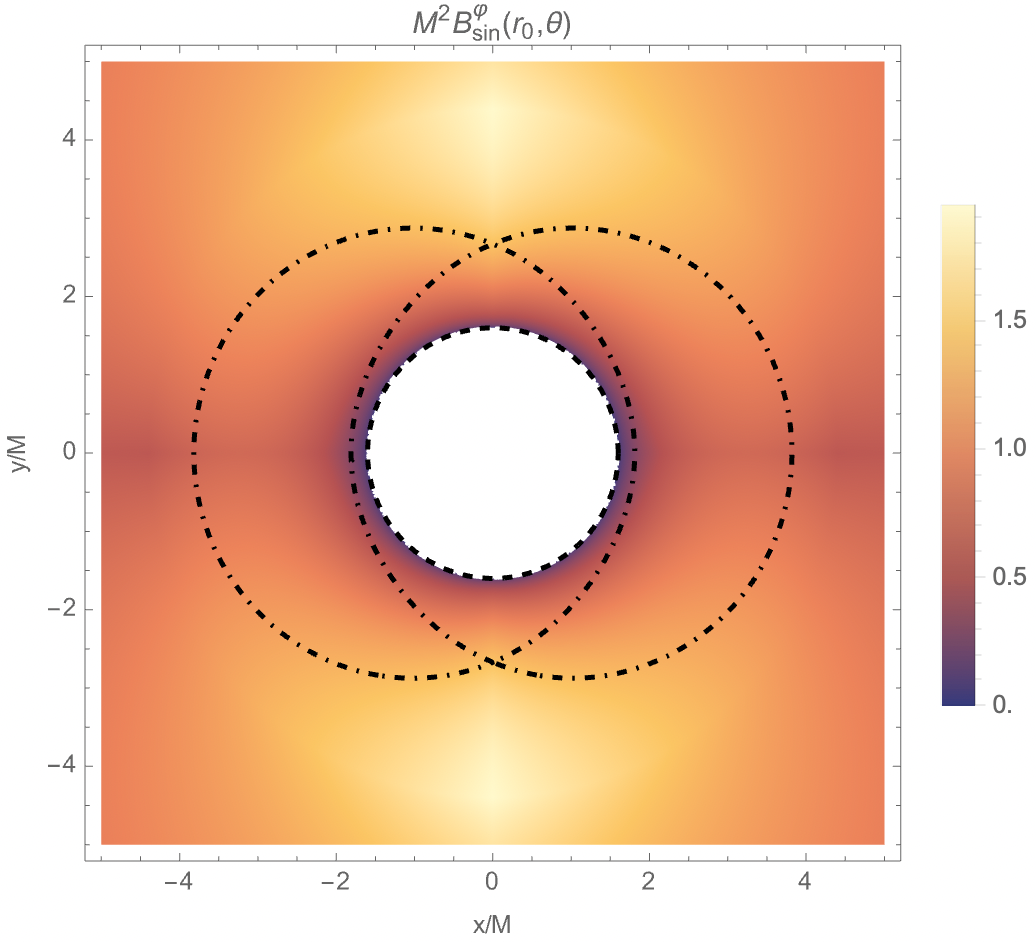}
\caption{The dimensionless quantities $M B^r_{(1)}$, $M^2 B^\theta_{(1)}$ and $M^2 B^\varphi_{(1)}$ of the induced magnetic field components are plotted as cosine part and sine part respectively. In all the figures we have set $a_0=0.8$ and $\alpha=0.2$. }
\label{Figures_B}
\end{figure}

The symmetries of these components are analyzed in Appendix \ref{Symmetries of the electric and magnetic fields}, consistent with the numerical results shown in FIG. \ref{Figures_E}-\ref{Figures_B}.

\begin{figure}[h]
\centering
  \includegraphics[width=0.3\textwidth]{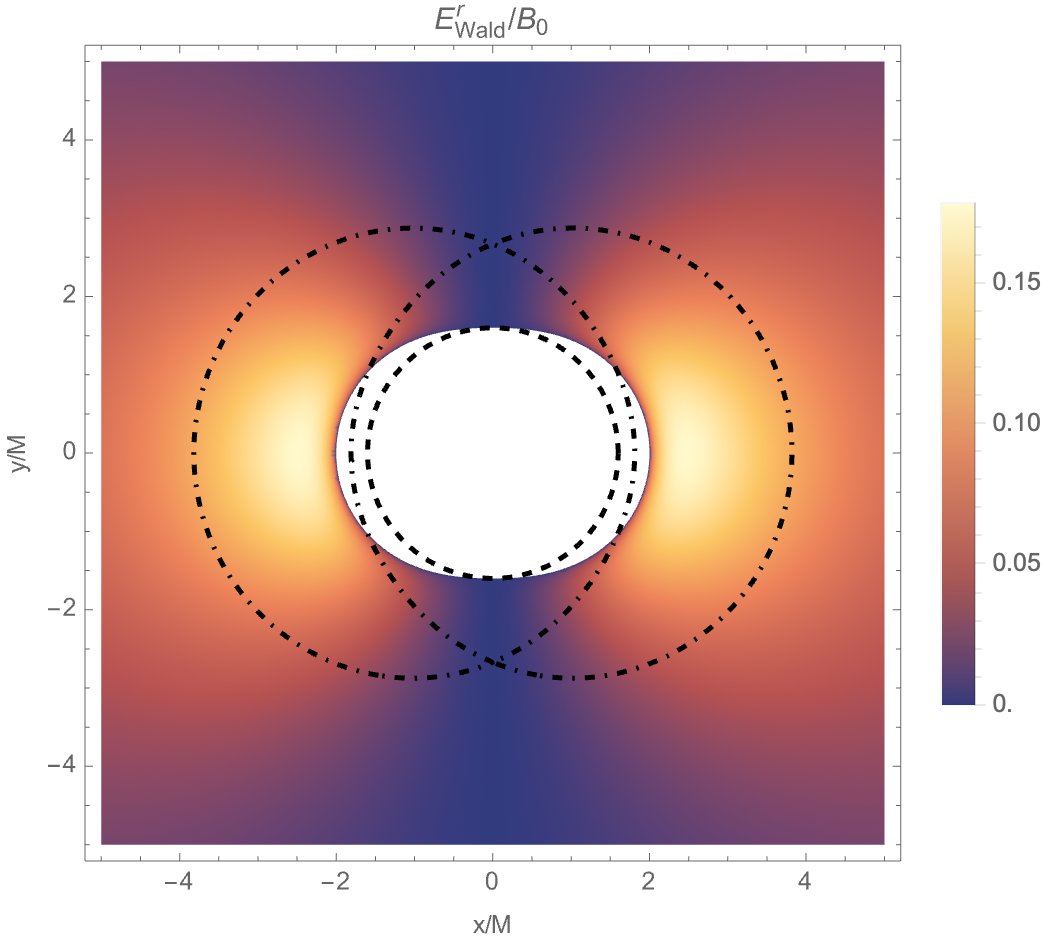}
  \includegraphics[width=0.3\textwidth]{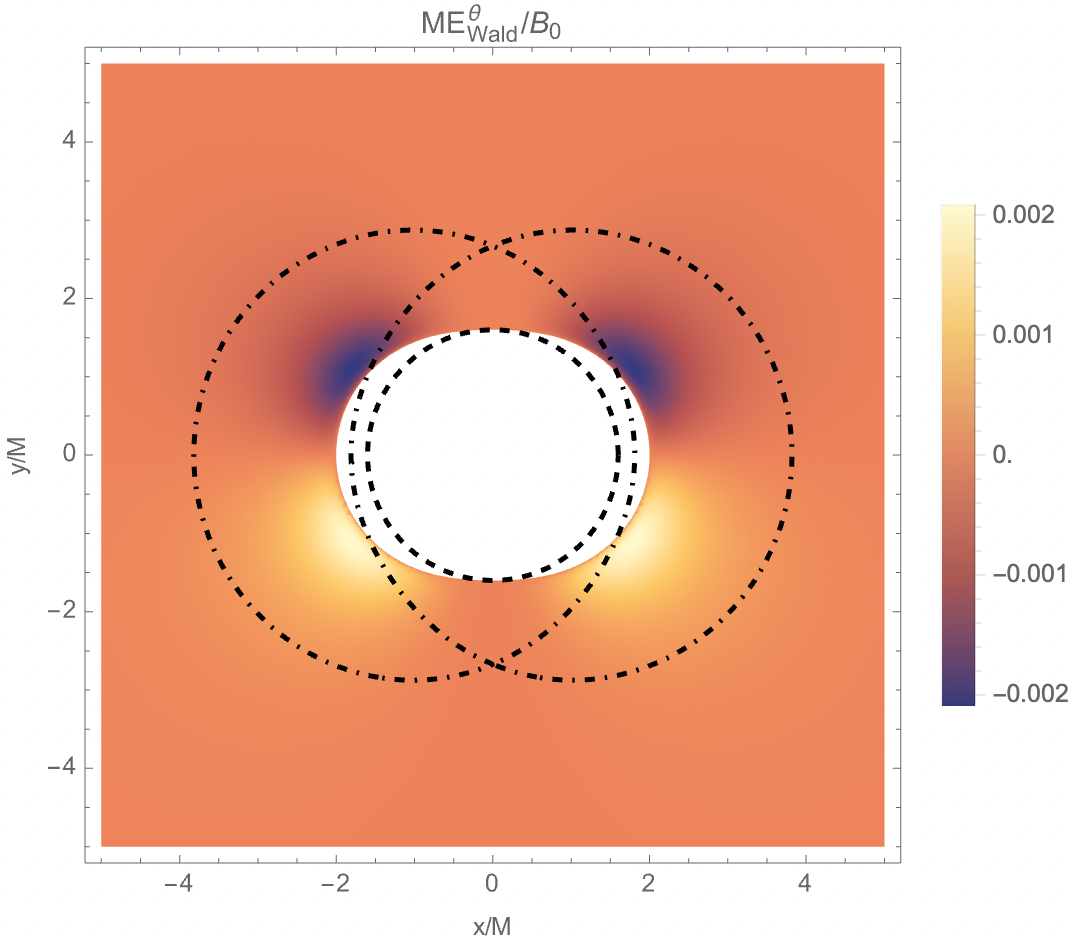}\\
  \includegraphics[width=0.3\textwidth]{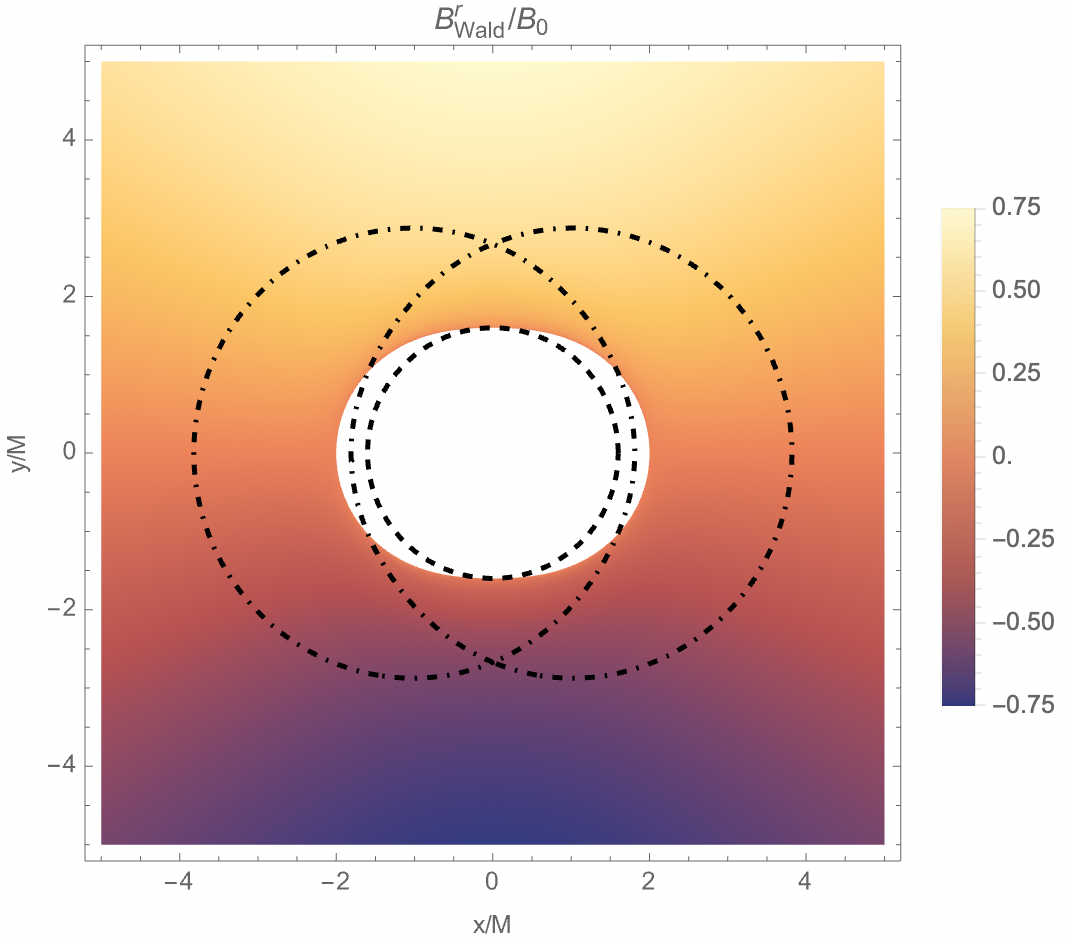}
  \includegraphics[width=0.3\textwidth]{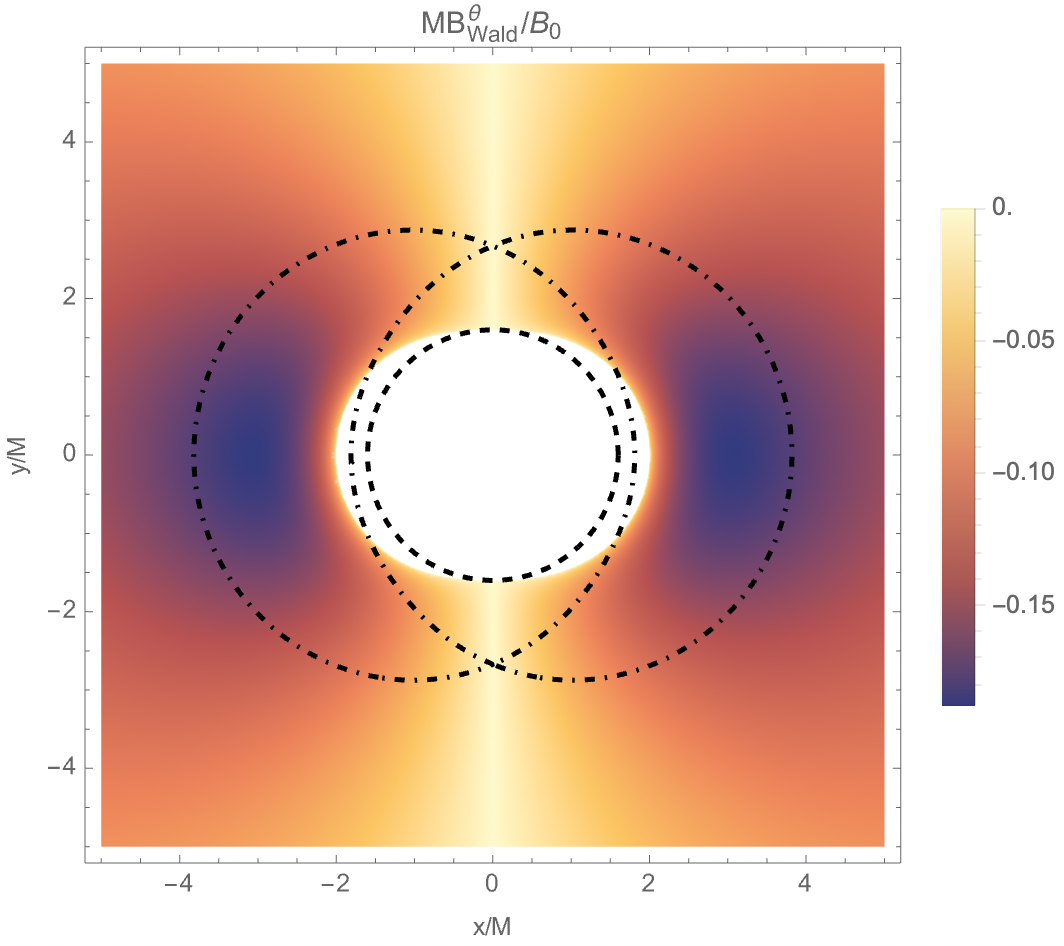}
\caption{The dimensionless components $E^r_{\rm Wald}/B_0$, $M E^\theta_{\rm Wald}/B_0$, $B^r_{\rm Wald}/B_0$, $M B^\theta_{\rm Wald}/B_0$ of the electric and magnetic fields are plotted for the background Wald case, where we have set $a_0=0.8$. }
\label{Figures_Wald}
\end{figure}

In FIG. \ref{Figures_E}-\ref{Figures_Wald}, we plot the components of induced electric and magnetic fields in cosine and sine sections respectively, in contrast with the background Wald case. In some panels with large scale like $100M$, we use the dotted circles to denote the radius $r_{\rm max}/M=2/\alpha^2$ of the maximum value of axion cloud, while in some panels with small scale like $5M$, we compare with the boundary of photon region (dotdashed curves) and event horizon (dashed circles). It is a good case to compare the distribution of the EMs (in field perspective) with the photon region (spherical photon orbits in particle perspective).

From the figures we can see that, the radial components $E^r_{\rm cos}$ and $E^r_{\rm sin}$ of the induced electric field both exhibit quadrupole structures with a sign reversal across the equatorial plane. In contrast, the radial component of the electric field in the background Wald solution is in dipole structure that is symmetric about the equatorial plane. While the $\theta$-components $E^\theta_{\rm cos}$ and $E^\theta_{\rm sin}$ of the induced electric field both exhibit even symmetry with respect to the equatorial plane, the $\theta$-component of the electric field in the background Wald solution shows a quadrupole pattern with odd symmetry. In addition, the $\varphi$-components $E^\varphi_{\rm cos}$ and $E^\varphi_{\rm sin}$ of the induced electric field possess odd symmetry about the equatorial plane. Conversely, the magnetic field components have opposite symmetries to those of the electric field components. Besides, we can observe some evident coincides between the distribution of the induced electric and magnetic fields and the boundary of photon region.

Note that both electric and magnetic fields of the background Wald case do not contain $\varphi$-components, and they only distribute outside the ergoregion, while the induced EM field can live within the ergoregion. This can be understood as ``no static observer is allowed inside the ergoregion".

Moreover, we are interested in the strengths of the induced electric and magnetic fields that
\begin{eqnarray}
    E^i_{(1)}E_i^{(1)}&=&\cos^2{\left(m \varphi -\omega t\right)}E^i_{\rm cos}E_i^{\rm cos}+\sin^2{\left(m \varphi -\omega t\right)}E^i_{\rm sin}E_i^{\rm sin}+2\cos{\left(m \varphi -\omega t\right)}\sin{\left(m \varphi -\omega t\right)}E^i_{\rm cos}E_i^{\rm sin}~\\
    &=&\cos{\left(2m \varphi -2\omega t\right)}\left(\frac{1}{2}E^i_{\rm cos}E_i^{\rm cos}-\frac{1}{2}E^i_{\rm sin}E_i^{\rm sin}\right)+\sin{\left(2m \varphi -2\omega t\right)}E^i_{\rm cos}E_i^{\rm sin}+\frac{1}{2}E^i_{\rm cos}E_i^{\rm cos}+\frac{1}{2}E^i_{\rm sin}E_i^{\rm sin} \notag 
\end{eqnarray}
which also works for $B^i_{(1)}B_i^{(1)}$. 

\begin{figure}[h]
\centering
  \includegraphics[width=0.3\textwidth]{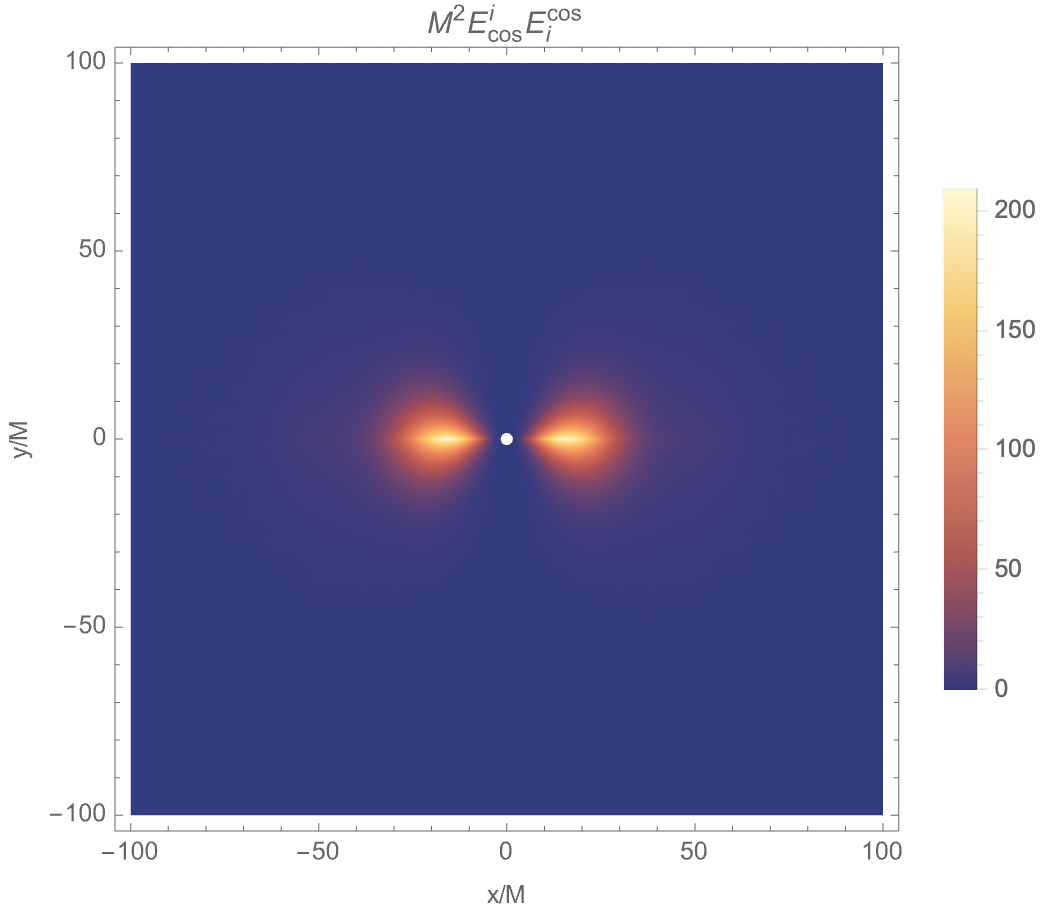}
  \includegraphics[width=0.3\textwidth]{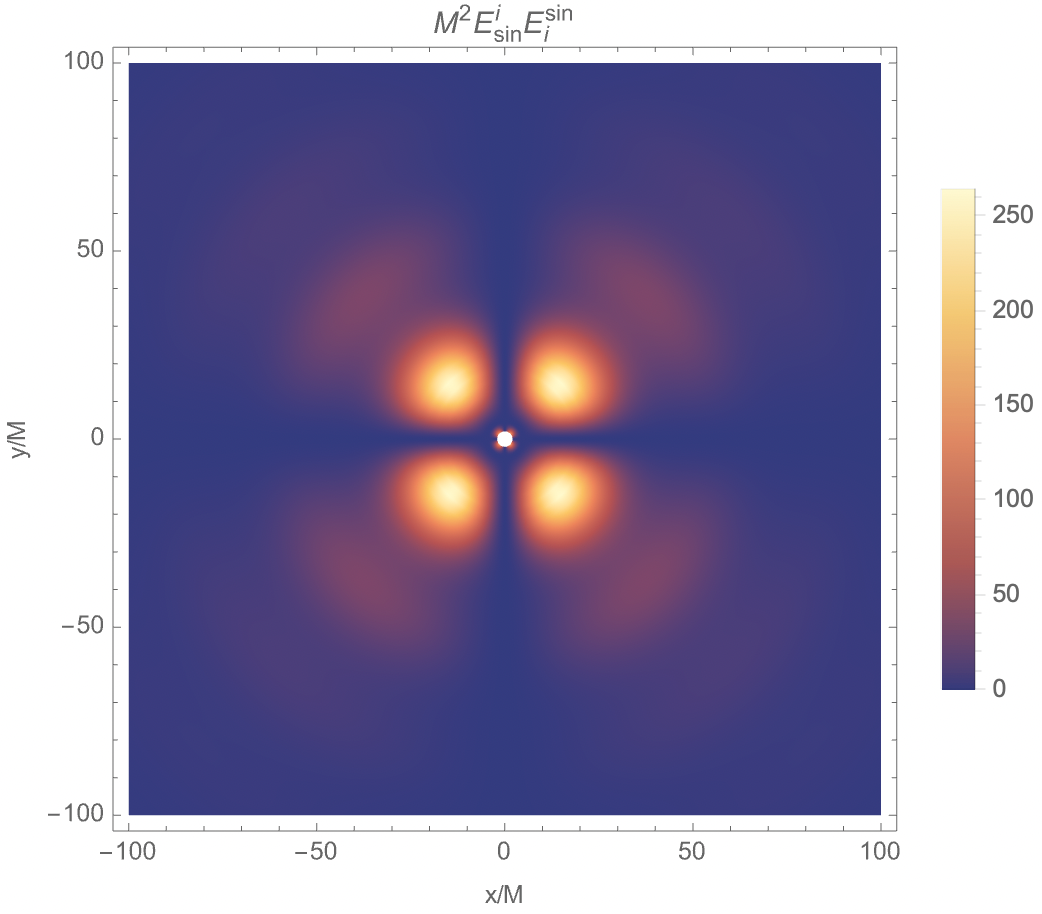}
  \includegraphics[width=0.3\textwidth]{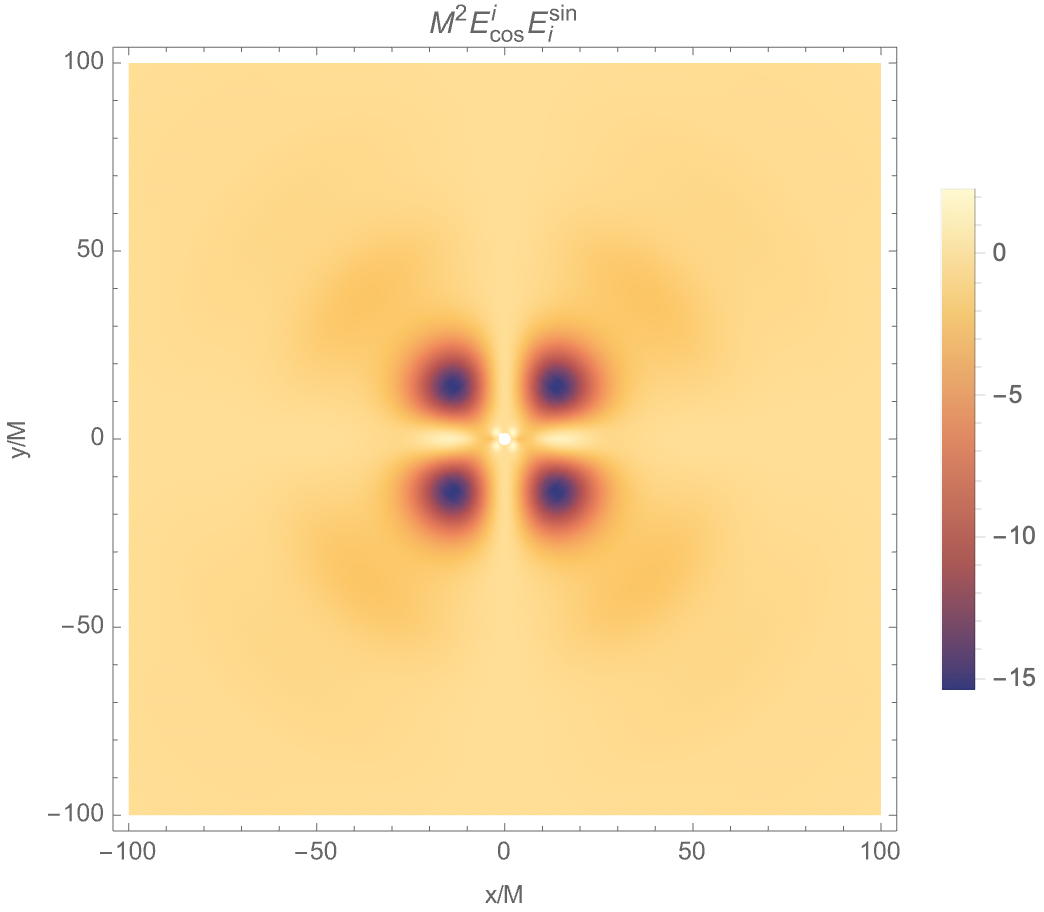}
  \includegraphics[width=0.3\textwidth]{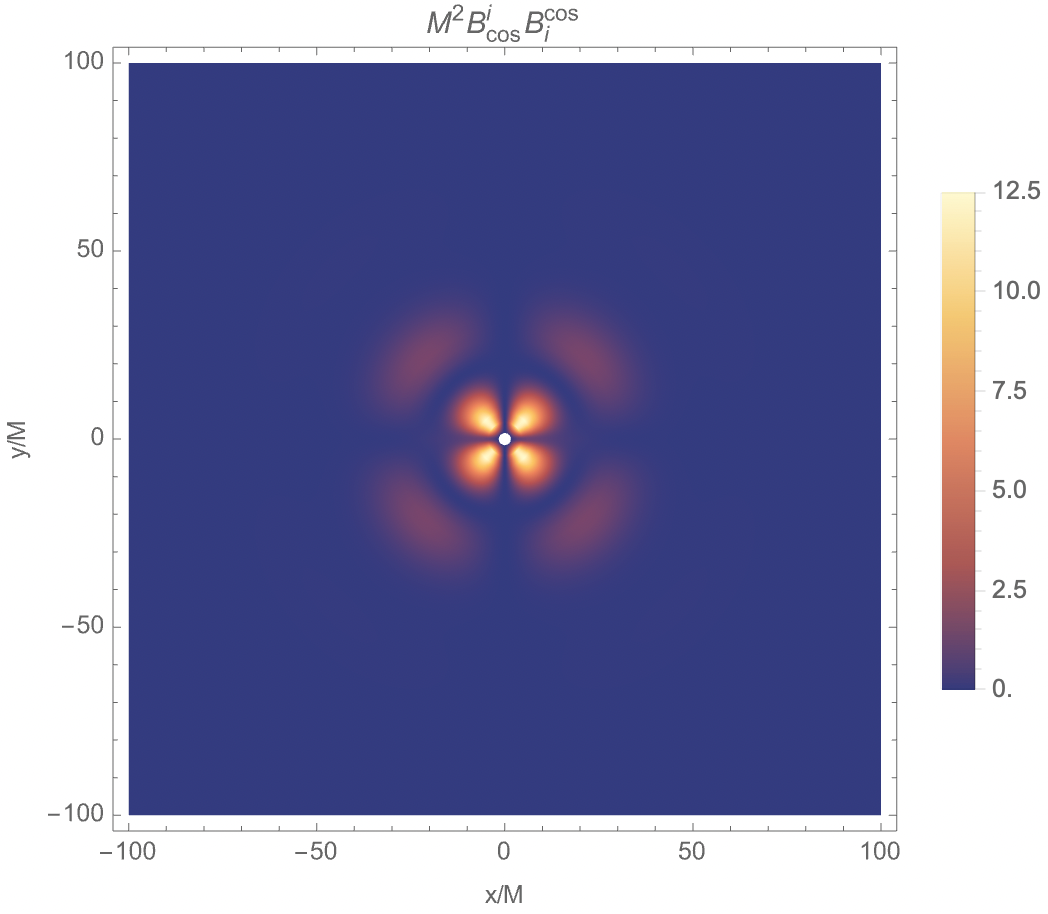}
  \includegraphics[width=0.3\textwidth]{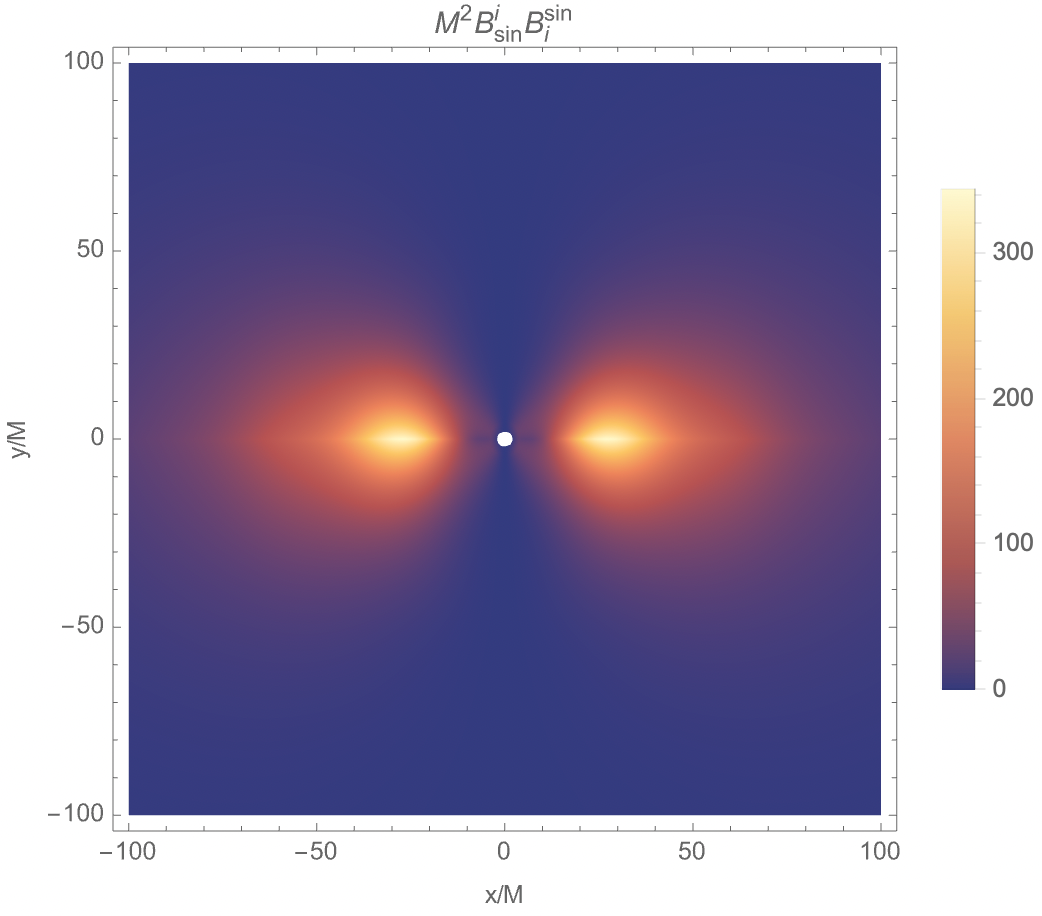}
   \includegraphics[width=0.3\textwidth]{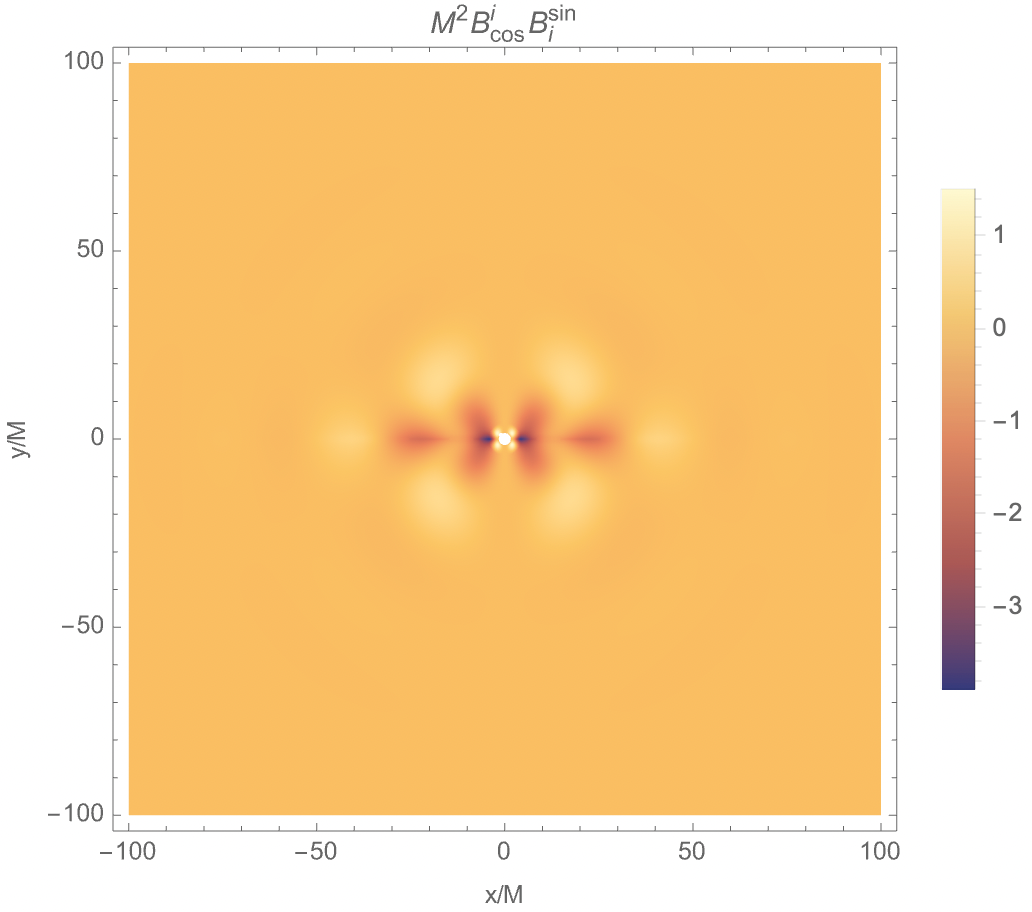}
\caption{The strengths of electric field and magnetic field $M^2 E^i_{\rm cos}E^{\rm cos}_i$, $M^2 E^i_{\rm sin}E^{\rm sin}_i$, $M^2 E^i_{\rm cos}E^{\rm sin}_i$, $M^2 B^i_{\rm cos}B^{\rm cos}_i$, $M^2 B^i_{\rm sin}B^{\rm sin}_i$, $M^2 B^i_{\rm cos}B^{\rm sin}_i$ are plotted with $a_0=0.8$ and $\alpha=0.2$.}
\label{Figures_EB2}
\end{figure}

We plot the strengths of the electric and magnetic fields in different sections in FIG. \ref{Figures_EB2}. It can be seen that the field strengths are mainly concentrated in $E^2_{\rm cos}$, $E^2_{\rm sin}$ and $B^2_{\rm sin}$, and that each panel exhibits a distinctive distribution pattern. As mentioned in~\cite{Teukolsky:1972my}, when sources are present, one can use the eigenfunctions of the angular equation to separate the master equation for $\varphi_0$ or $\varphi_2$ (their source-free master equations are separable) by expanding the source as $T=\Sigma G(r)_s S_{lm}(\theta)e^{im\varphi}e^{-i\omega t}$. Recall that there are only two real degrees of freedom; once $\varphi_0$ or $\varphi_2$ is determined, the other two complex quantities are fixed. Therefore, these featured patterns should come from the decomposition of the source. In our case, the source is the axion cloud, but the expression of $T$ is too complicate to expand term by term.

As it is well known, superradiance occurs only for massive fields, where the mass term appears in the effective potential as an asymptotic constant at spatial infinity, providing a naturally reflective boundary that traps waves with lower oscillation frequencies. In addition, the mass term localizes the field around the black hole and allows it to sustainably dissipate near the event horizon. The dissipation is a crucial ingredient for superradiance. In fact, even for a rotating compact object without a horizon, superradiance can occur through other dissipation mechanisms, such as friction~\cite{Brito:2015oca}. 

In the case of axion-photon coupling, the antisymmetry of  $\epsilon^{\mu\nu\rho\sigma}$ prevents the photons from acquiring an effective mass term for the photons; this is sometimes referred to as a ``tachyonic-like instability"~\cite{Ikeda:2018nhb}. It has also been noted that a time-dependent axion is crucial for triggering this instability when $k_a \phi_0$ exceeds a critical value. Furthermore, as mentioned earlier, in the presence of an external magnetic field, a flat spacetime analysis can give the mixing matrices for both the axion-photon coupling and the dark photon-photon coupling cases~\cite{Cannizzaro:2024hdg}
\begin{equation}
    \mathcal{M}_{a\gamma}=
    \begin{pmatrix}
    -\omega_p^2/2\omega & B_y k_a \\
    B_y k_a & -\mu_a^2/2\omega 
    \end{pmatrix}, \quad \mathcal{M}_{\gamma \gamma'}=\frac{1}{2\omega}
    \begin{pmatrix}
    -\omega_p^2 & \sin{\chi_0}\mu^2_{\gamma'} \\
    \sin{\chi_0}\mu^2_{\gamma'} & -\mu_{\gamma'}^2 
    \end{pmatrix},
\end{equation}
where the presence of a magnetic field enables the axion-photon coupling to effectively act as a ``mass term" in the mixing, whereas the dark photon-photon mixing matrix does not require an external magnetic field. Moreover, the flat spacetime analysis shows that in a homogeneous EM background, the dispersion relation is modified by the axion-photon coupling~\cite{Boskovic:2018lkj}
\begin{equation}
    \omega^2=\mathfrak{p}^2 +\frac{1}{2}\mu^2\pm \frac{1}{2}\sqrt{\mu^4+16k_a^2 E_z^2 \mathfrak{p}^2}~
\end{equation}
where a sufficiently large electric field leads to $\omega^2<0$, corresponding to a positive imaginary part of $\omega$ and thus a growing mode (instability). This is very similar to the superradiance case. Particularly the axions and photons share the same frequency, although the instability is driven by a stronger electric field, unlike our case.

\section{Conclusions}
\label{Conclusions}

In this work we studied the EM photon cloud sourced by an axion cloud through the axion-photon coupling around a Kerr black hole in the presence of a background EM field. At first order in the coupling constant, we obtained an oscillating EM photon cloud with the same frequency as the axion cloud, indicating that it can exist even for very small couplings, with no threshold for either the coupling strength or the EM background field amplitude. 

More intriguingly, this induced EM photon cloud grows exponentially in accordance with the axion cloud when the superradiant condition for the axion field is satisfied within the perturbative framework. This instability is essentially inherited from the axion cloud, distinct from the instability that originates in the homogeneous part of the equation via the Mathieu equation, and also from that arising in the inhomogeneous (sourced) part of the EM field equation, where a sufficiently strong electric field is required.

The growth and spatial distribution of the induced EM field are thus primarily governed by the axion cloud and the EM background field, though with various featured patterns. Both the EM background and the axion cloud serve as sources driving the growth of the induced EM photon cloud, which is mainly concentrated near the black hole while dissipating sustainably from the axion cloud to the photon region and finally toward the event horizon. As shown in the figures, the induced EM field does not peak at the location of the maximum axion density; rather, its main distribution lies within the dotted circles. In other words, the densest regions of the photon cloud do not coincide with the radius of maximum axion oscillation (same as $r_{\rm max}$) but are located inside it. 

In addition, we investigated the induced electric and magnetic fields respectively, and found that their components exhibit markedly different symmetries compared to those of the background EM field. As a result, the induced photon cloud forms an unstable bound configuration that emits EM waves to spatial infinity while being replenished by the axion cloud, providing a potential observational signature of both the presence of an axion cloud and axion-photon coupling. While a detailed comparison with black hole polarization images is beyond the scope of this work, the evolving EM structures induced by axion–photon coupling could, in principle, indirectly imprint subtle signatures in polarization observables—such as deviations from standard GRMHD polarization patterns or circular polarization—through their influence on plasma dynamics and radiative processes.

Furthermore, we analyzed the symmetries of most relevant quantities, which are consistent with the numerically results shown in the figures. Besides, in the near-horizon region, we compared the distribution of the induced EM field with the boundary of photon region and found partial overlap, as expected.

\acknowledgments

We thank Masahide Yamaguchi, Jin Sun, Yuhang Zhu, Minxi He and Cheng-Yong Zhang for beneficial discussions. We thank the editor and the referees for their constructive comments and suggestions. We also thank Enrico Cannizzaro for his comments, which improved our interpretations. Z.-Y. Tang is supported by IBS under the project code IBS-R018-D3.

\appendix{}

\section{Asymptotic behaviors of $f_{01M}$, $f_{11M}$ and $f_{\rm 21M}$}
\label{Asymptotic behaviors}

With expansions and the regular condition at poles $\theta=0,\pi$, the asymptotic behaviors at the boundaries $r_0=r_{\rm h0}\equiv 1+\sqrt{1-a_0^2}$, $r_0 \to \infty$, $\theta=0,\pi$ are as follows
\begin{eqnarray}
    &&f_{\rm 01M}\left(r_{\rm h0},\theta\right)=\frac{2 \sqrt{2} a_0 r_{h0}^3 e^{-\alpha^2 r_{\rm h0}/2} \sin{\theta} \sin {2 \theta} }{\left(2 \alpha r_{h0}-a_0-2i \sqrt{1-a_0^2}\right) \left(r_{\rm h0}^2+a_0^2 \cos ^2{\theta }\right)^2}~,\\
    &&f_{\rm 11M}\left(r_{\rm h0},\theta\right)=\frac{2 i \sqrt{1-a_0^2} r_{h0}^3 e^{-\alpha^2 r_{\rm h0}/2} \sin {2 \theta} }{\left(a_0-2 \alpha r_{h0}\right) \left(r_{\rm h0}^2+a_0^2 \cos ^2{\theta }\right)^2}~,\\
    &&f_{\rm 21M}\left(r_{\rm h0},\theta\right)=0~,\\
    &&f_{\rm 01M}\left(r_0 \to \infty,\theta\right)=-\frac{3 i \sin ^2{\theta} e^{-\alpha^2 r_0/2}}{\sqrt{2} \alpha}+\mathcal{O}\left(r_0^{-1} e^{-\alpha^2 r_0/2}\right)~,\\
    &&f_{\rm 11M}\left(r_0 \to \infty,\theta\right)=-\frac{(\alpha+3 i) \sin {2 \theta}  e^{-\alpha^2 r_0/2}}{4 \alpha}+\mathcal{O}\left(r_0^{-1} e^{-\alpha^2 r_0/2}\right)~,\\
    &&f_{\rm 21M}\left(r_0 \to \infty,\theta\right)=-\frac{3 i \sin ^2{\theta } e^{-\alpha^2 r_0/2}}{2 \sqrt{2} \alpha}+\mathcal{O}\left(r_0^{-1} e^{-\alpha^2 r_0/2}\right)~,\\
    &&f_{\rm 01M}\left(r_0,0\right)=0, \quad
    f_{\rm 11M}\left(r_0,0\right)=0, \quad
    f_{\rm 21M}\left(r_0,0\right)=0, \\
    &&f_{\rm 01M}\left(r_0,\pi\right)=0, \quad
    f_{\rm 11M}\left(r_0,\pi\right)=0, \quad
    f_{\rm 21M}\left(r_0,\pi\right)=0~,
\end{eqnarray}
where the outgoing part $f_{\rm 21M}\left(r_0,\theta\right)$ just vanishes at the horizon in accord with the fact that there is no outgoing radiation at the event horizon. It should also result from the fact that $\varphi_2$ is with negative spin weight which can not retrograde against the rotation at event horizon. 

In addition, it is clear that the asymptotic behaviors for $f_{\rm 01M}$ and $f_{\rm 11M}$ at event horizon $r_{\rm h0}$ are only valid when $\alpha \neq \frac{a_0}{2r_{\rm h0}}$, which is exactly the critical condition for whether superradiance occurs. In this critical case with $\alpha=\frac{a_0}{2r_{\rm h0}}$, the imaginary part $\omega_I$ disappears in the original field equation for the axion field, so that the whole frequency is not complex and the original boundary condition at event horizon is not valid, in which case another regular condition at origin should be applied. Here we do not go further on this case. 

\section{Components of $F_{\mu\nu}^{(1)}{}^*F^{\mu\nu}_{(1)}$ and $F_{\mu\nu}^{(1)}F^{\mu\nu}_{(1)}$ }
\label{FF}

It is well known that the term $F_{\mu\nu}{}^*F^{\mu\nu}$ violates parity and couples to the pseudo-scalar axion field, making the whole term $\phi F_{\mu\nu}{}^*F^{\mu\nu}$ parity invariant and leading to CP violation.

The components of the parity violation term $F_{\mu\nu}^{(1)}{}^*F^{\mu\nu}_{(1)}$ 
\begin{eqnarray}
    F_{\mu\nu}^{\rm cos}{}^*F^{\mu\nu}_{\rm cos}&=&8 \left[({\rm Im}[f_{01}] +{\rm Im}[f_{02}] ) ({\rm Re}[f_{21}] +{\rm Re}[f_{22}] )+({\rm Re}[f_{01}] +{\rm Re}[f_{02}] ) ({\rm Im}[f_{21}] +{\rm Im}[f_{22}] )\right.\notag\\
    &&\left.-2 ({\rm Im}[f_{11}] +{\rm Im}[f_{12}] ) ({\rm Re}[f_{11}] +{\rm Re}[f_{12}] )\right]~,\\
    F_{\mu\nu}^{\rm sin}{}^*F^{\mu\nu}_{\rm sin}&=&-8 ({\rm Im}[f_{01}] -{\rm Im}[f_{02}] ) ({\rm Re}[f_{21}] -{\rm Re}[f_{22}] )-8 ({\rm Re}[f_{01}] -{\rm Re}[f_{02}] ) ({\rm Im}[f_{21}] -{\rm Im}[f_{22}] ) \notag\\
    &&+16 ({\rm Im}[f_{11}] -{\rm Im}[f_{12}] ) ({\rm Re}[f_{11}] -{\rm Re}[f_{12}] )~,\\
    F_{\mu\nu}^{\rm cos}{}^*F^{\mu\nu}_{\rm sin}&=&8 \left[-{\rm Im}[f_{01}]  {\rm Im}[f_{21}] +{\rm Re}[f_{01}]  {\rm Re}[f_{21}] +{\rm Im}[f_{02}]  {\rm Im}[f_{22}] -{\rm Re}[f_{02}]  {\rm Re}[f_{22}] \right.\notag\\
    &&\left.+{\rm Im}[f_{11}] ^2-{\rm Re}[f_{11}] ^2-{\rm Im}[f_{12}] ^2+{\rm Re}[f_{12}] ^2\right]~,
\end{eqnarray}
all exhibit a sign change for $\theta \to \pi -\theta$, in agreement with the results in FIG. \ref{Figures_FFD}. Combined with no sign change for $\varphi \to \varphi +\pi$, it shows a parity violation $P=-1$ for $F_{\mu\nu}^{(1)}{}^*F^{\mu\nu}_{(1)}$. 

While the components of the contraction of the EM field strength tensor $F_{\mu\nu}^{(1)}F^{\mu\nu}_{(1)}$
\begin{eqnarray}
    F_{\mu\nu}^{\rm cos}F^{\mu\nu}_{\rm cos}&=&8 \left[-({\rm Im}[f_{01}] +{\rm Im}[f_{02}] ) ({\rm Im}[f_{21}] +{\rm Im}[f_{22}] )+({\rm Re}[f_{01}] +{\rm Re}[f_{02}] ) ({\rm Re}[f_{21}] +{\rm Re}[f_{22}] )\right.~\notag\\
    &&\left.+({\rm Im}[f_{11}] +{\rm Im}[f_{12}] )^2-({\rm Re}[f_{11}] +{\rm Re}[f_{12}] )^2\right]~,\\
    F_{\mu\nu}^{\rm sin}F^{\mu\nu}_{\rm sin}&=&8 \left[({\rm Im}[f_{01}] -{\rm Im}[f_{02}] ) ({\rm Im}[f_{21}] -{\rm Im}[f_{22}] )-({\rm Re}[f_{01}] -{\rm Re}[f_{02}] ) ({\rm Re}[f_{21}] -{\rm Re}[f_{22}] )\right.\notag\\
    &&\left.-({\rm Im}[f_{11}] -{\rm Im}[f_{12}] )^2+({\rm Re}[f_{11}] -{\rm Re}[f_{12}] )^2\right]~,\\
    F_{\mu\nu}^{\rm cos}F^{\mu\nu}_{\rm sin}&=&8 \left[-{\rm Im}[f_{01}]  {\rm Re}[f_{21}] -{\rm Re}[f_{01}]  {\rm Im}[f_{21}] +{\rm Im}[f_{02}]  {\rm Re}[f_{22}] +{\rm Re}[f_{02}]  {\rm Im}[f_{22}]\right. \notag\\
    &&\left.+2 {\rm Im}[f_{11}]  {\rm Re}[f_{11}] -2 {\rm Im}[f_{12}]  {\rm Re}[f_{12}] \right]~,
\end{eqnarray}
all keep a parity symmetry $P=+1$ under $\theta \to \pi -\theta$, in agreement with the results in FIG. \ref{Figures_FF}. Also there is no sign change under $\varphi \to \varphi +\pi$, hence the whole $F_{\mu\nu}^{(1)}F^{\mu\nu}_{(1)}$ keeps parity symmetry $P=+1$. 

We know that the null conditions for EM fields are $F_{\mu\nu}F^{\mu\nu}=0$ and $F_{\mu\nu}{}^*F^{\mu\nu}=0$, which is equivalent to $\varphi_0 \varphi_2-\varphi_1^2=0$. Here with the axion-photon coupling, it shows obviously that the null conditions are not satisfied anymore. 

\section{Symmetries of the electric and magnetic fields}
\label{Symmetries of the electric and magnetic fields}

For the electric and magnetic fields respectively, applying the relations (\ref{relationf0})-(\ref{relationf2}), we can obtain some symmetries for the components in cosine and sine parts
\begin{eqnarray}
    &&E^r_{\rm cos/sin}\left(r,\pi-\theta\right)=-E^r_{\rm cos/sin}\left(r,\theta\right)~,\quad B^r_{\rm cos/sin}\left(r,\pi-\theta\right)=B^r_{\rm cos/sin}\left(r,\theta\right)~, \\
    &&E^\theta_{\rm cos/sin}\left(r,\pi-\theta\right)=E^\theta_{\rm cos/sin}\left(r,\theta\right)~,\quad B^\theta_{\rm cos/sin}\left(r,\pi-\theta\right)=-B^\theta_{\rm cos/sin}\left(r,\theta\right)~, \\
    &&E^\varphi_{\rm cos/sin}\left(r,\pi-\theta\right)=-E^\varphi_{\rm cos/sin}\left(r,\theta\right)~,\quad B^\varphi_{\rm cos/sin}\left(r,\pi-\theta\right)=B^\varphi_{\rm cos/sin}\left(r,\theta\right)~, 
\end{eqnarray}
which can be clearly seen in FIG. \ref{Figures_E}-\ref{Figures_B}. We further have
\begin{eqnarray}
    &&E^r_{(1)}\left(t,r,\pi-\theta,\varphi+\pi\right)=E^r_{(1)}\left(t,r,\theta,\varphi\right)~,\quad B^r_{(1)}\left(t,r,\pi-\theta,\varphi+\pi\right)=-B^r_{(1)}\left(t,r,\theta,\varphi\right)~, \\
    &&E^\theta_{(1)}\left(t,r,\pi-\theta,\varphi+\pi\right)=-E^\theta_{(1)}\left(t,r,\theta,\varphi\right)~, \quad B^\theta_{(1)}\left(t,r,\pi-\theta,\varphi+\pi\right)=B^\theta_{(1)}\left(t,r,\theta,\varphi\right)~, \\
    &&E^\varphi_{(1)}\left(t,r,\pi-\theta,\varphi+\pi\right)=E^\varphi_{(1)}\left(t,r,\theta,\varphi\right)~, \quad B^\varphi_{(1)}\left(t,r,\pi-\theta,\varphi+\pi\right)=-B^\varphi_{(1)}\left(t,r,\theta,\varphi\right)~, 
\end{eqnarray}
and the background EM field transforms as
\begin{eqnarray}
    &&E^r_{(0)}\left(t,r,\pi-\theta,\varphi+\pi\right)=E^r_{(0)}\left(t,r,\theta,\varphi\right)~, \quad B^r_{(0)}\left(t,r,\pi-\theta,\varphi+\pi\right)=-B^r_{(0)}\left(t,r,\theta,\varphi\right)~, \\
    &&E^\theta_{(0)}\left(t,r,\pi-\theta,\varphi+\pi\right)=-E^\theta_{(0)}\left(t,r,\theta,\varphi\right)~, \quad B^\theta_{(0)}\left(t,r,\pi-\theta,\varphi+\pi\right)=B^\theta_{(0)}\left(t,r,\theta,\varphi\right)~.
\end{eqnarray}

These transformations are equivalent to $\overset{\rightharpoonup}{E}\rightarrow -\overset{\rightharpoonup}{E}$ and $\overset{\rightharpoonup}{B}\rightarrow \overset{\rightharpoonup}{B}$ under $\overset{\rightharpoonup}{r}\rightarrow -\overset{\rightharpoonup}{r}$. Therefore the induced electric field still keep a parity symmetry with $P=-1$ as a vector, and the induced magnetic field keeps the parity symmetry with $P=+1$ as a pseudo-vector, same as the EM background.

\end{document}